\documentclass[onecollarge]{svjour2}       
\usepackage{graphics}
\usepackage{graphicx}
\usepackage{alltt}
\usepackage{wasysym}
\usepackage{amssymb}
\usepackage{color}
\usepackage{epsfig}
\usepackage[normalem]{ulem}
\begin{document}
\title{Tidal synchronization of close-in satellites and exoplanets. III. Tidal dissipation revisited and application to Enceladus.}
\subtitle{\today}
\author{H.A. Folonier \and S. Ferraz-Mello \and E. Andrade-Ines}
\institute{H.A. Folonier, S. Ferraz-Mello and E.Andrade-Ines \at \emph{Instituto de Astronomia Geof\'isica e Ci\^encias
Atmosf\'ericas, Universidade de S\~ao Paulo, Brasil} \\
\email{folonier@usp.br}, sylvio@iag.usp.br and eandrade.ines@gmail.com\\}
\titlerunning{Tidal dissipation}
\maketitle
\def\hig{\textbf}
\begin{abstract}
	This paper deals with a new formulation of the creep tide theory (Ferraz-Mello, Cel. Mech. Dyn. Astron. {\bf 116}, 109, 2013 $-$ Paper I) and with the tidal 
	dissipation predicted by the theory in the case of stiff bodies whose rotation is not synchronous but is oscillating around the synchronous state with 
	a period equal to the orbital period. We show that the tidally forced libration influences the amount of energy dissipated in the body and the average 
	perturbation of the orbital elements. This influence depends on the libration amplitude and is 
	generally neglected in the study of planetary satellites. However, they may be responsible for a 27 percent increase in the dissipation of Enceladus. The 
	relaxation factor necessary to explain the observed dissipation of Enceladus ($\gamma=1.2-3.8\times 10^{-7}\ {\rm s}^{-1}$) has the expected order of 
	magnitude for planetary satellites and corresponds to the viscosity $0.6-1.9 \times 10^{14}$ Pa s, which is in reasonable agreement with the value recently 
	estimated by Efroimsky (2018) ($0.24 \times 10^{14}$ Pa s) and with the value adopted by Roberts and Nimmo (2008) for the viscosity of the ice shell 
	($10^{13}-10^{14}$ Pa s). For comparison purposes, the results are extended also to the case of Mimas and are consistent with the negligible dissipation and 
	the absence of observed tectonic activity. The corrections of some mistakes and typos of paper II (Ferraz-Mello, Cel. Mech. Dyn. Astron. {\bf 122}, 359, 
	2015) are included at the end of the paper.
	
\end{abstract}
\def\beq{\begin{equation}}
\def\endeq{\end{equation}}
\def\begdi{\begin{displaymath}}
\def\enddi{\end{displaymath}}
\def\ep{\varepsilon}
\def\epp{\varepsilon^\prime}
\def\eppp{\varepsilon^{\prime\prime}}
\def\App{A^\prime}
\def\Appp{A^{\prime\prime}}
\def\bpp{\beta^\prime}
\def\bppp{\beta^{\prime\prime}}
\def\Cpp{C^\prime}
\def\Cppp{C^{\prime\prime}}
\def\CppQ{C^{\prime 2}}
\def\CpppQ{C^{\prime\prime 2}}
\def\aprmenor{\;\buildrel{<}\over{\sim}\;}              
\def\aprmaior{\;\buildrel\hbox{$>$}\over{\sim}\;}       
\def\defeq{\;\buildrel\hbox{\small def}\over{\,=}\;}    
\def\speq{\hspace{1mm} = \hspace{1mm}}                  
\def\FRH{ Ferraz-Mello et al (2008) }
\def\SFM{paper I }
\def\Z{\mathbb{Z}}
\def\D{\rm{d}}
\def\hilight{\textbf}
\def\half{\frac{1}{2}}

\section{Introduction}\label{sec01}

The calculus of the energy dissipation inside a stiff body is generally done by estimating the dissipation resulting from the action of the primary tidal
force in deforming the planet (Kopal, 1963; Kaula, 1963, 1964; Peale and Cassen, 1978; Segatz et al., 1988; Wisdom, 2008; Shoji et al., 2013; Frouard and 
Efroimsky, 2017). Such approach involves the choice of the physical model of the forces acting on the body at the microscopic level and of the dissipation 
parameters inside the body. An indirect approach was discussed by Kaula (1964; p.677) in which the bulk dissipation is calculated from the estimation of the 
total mechanical energy lost by the system. This approach was used by Yoder and Peale (1981) to estimate the tidal energy dissipated in a synchronous 
satellite, by Lissauer et al. (1984) to study the melting of Enceladus, by Ferraz-Mello et al. (2008) and Ferraz-Mello (2013) to estimate the energy 
dissipated in the frame of Darwin's and creep tide theories, respectively, and by Correia et al. (2014) in the frame of a Maxwell model.

In this paper, we revisit the indirect approach to evaluate the bulk loss of mechanical energy by the system. This approach has the merit of its simplicity. 
If the companion body (the body responsible for the tide raised in the stiff body) is considered as a mass point, the energy tidally dissipated in 
the primary body (the stiff body under consideration) may only take origin in its rotation and in the orbit of the system. The secular variations of the 
semi-major axis and of the rotation of the body are the two gauges allowing us to evaluate the mechanical energy lost by the system. No other non-primeval 
source exists able to continuously add energy to the system. We thus consider the energy exchanged with the orbit due to the direct attraction of the two 
bodies, the stored rotational energy in the primary body. Several minor contributions, considered for the sake of completeness, are shown to be negligible. 

We remind that the only assumption of the adopted tide theory is that self-gravitation and tidal stress permanently adjust the surface of the body to an 
equilibrium surface with speed given by the Newtonian creep law. The adjustment is ruled by an approximate solution of the Navier-Stokes equation for the 
flow of matter in the immediate neighborhood of the equilibrium surface of the body. No constitutive equation linking strain and stress is introduced at any 
point in the creep tide theory. All developments to reach the conclusion are the solution of the creep differential equation and the use of classical 
Physics to compute the force and torque acting on the companion due to the tidal deformation of the primary. The observed dissipation law results directly 
from the above described first principles of physics, with approximations, but no additional ad-hoc hypotheses. However, the integration of the basic 
equation of the creep tide theory in papers I and II assumes that the rotation of the body and the Keplerian elements of the orbit do not show
significant variations in one orbital period. In the case of stiff bodies with a nearly synchronous rotation, however, it has been shown that the rotation 
of the bodies is not damped to a stationary value (as gaseous bodies) but is rather driven to a periodic attractor with the same period as the 
orbital motion and amplitude proportional to the orbital eccentricity (see Correia et al., 2014; Ferraz-Mello, 2015; Folonier, 2016)\footnote{This forced 
libration is related to the asymmetries of the tides raised on the body; it is different from the asymmetries resulting from the assumption of a permanent 
triaxiality of the body (see Frouard and Efroimsky, 2017).}. Consequently, in such case, this physical libration needs to be considered in the 
integration of the basic differential equation of the creep. In this way, we propose a new model for the classical theory, assuming that the shape of the 
tidally deformed body may be approximated by a triaxial ellipsoid, where its flattenings and orientation are unknown functions of the time (this idea is 
supported by the analytical solutions of the creep equation given in papers I and II). Then, the creep equation allows us to 
find the differential equations that describe the time evolution of this ellipsoidal bulge and its orientation, resulting in a very simpler and compact 
approach.

The results are applied to Enceladus and Mimas. These satellites are selected for this study because of the amount of observational information available. 
We know from Cassini's observations the second-degree components of the gravitational field (Iess et al., 2014). We also know that the crust of Enceladus 
presents a forced libration of $0.120 \pm 0.014$ degrees (Thomas et al. 2016) and that a huge quantity of heat flows from the satellite ($5-16$ GW cf Howett 
et al. 2011; Spencer et al. 2013; Le Gall et al. 2017). In contrast, Mimas presents a larger forced libration of $0.838 \pm 0.002$ degrees (Tajeddine et al., 
2014) and the absence of current tectonic activity is evidence of a small dissipation (smaller than 1 GW). In addition, in both cases the three radii of 
the best ellipsoid representing the satellite surface are known (see Archinal et al., 2018).

The presence of the physical libration affects the dissipation increasing it but the increase is not so important as to affect the order of magnitude of the 
dissipation and certainly not so important as some preliminary results of this investigation seemed to indicate. The amount of dissipation observed by 
Cassini corresponds to a relaxation factor in the range $\gamma = 1.2-3.8 \times 10^{-7}$ s$^{-1}$ for Enceladus and $\gamma \sim 10^{-9}\ {\rm s}^{-1}$ for 
Mimas. The difference between these values is consistent with the fact that the gravitational acceleration at the surface and the density are both much 
larger in Enceladus than in Mimas. On the other hand, these values of relaxation factors indicate a viscosity $\eta=0.6-1.9 \times 10^{14}$ Pa s for 
Enceladus and $\eta\sim 10^{16}$ Pa s for Mimas. This estimation for the Enceladus viscosity has the same order as the value recently estimated by 
Efroimsky (2018) ($0.24 \times 10^{14}$ Pa s) and as the value adopted by Roberts and Nimmo (2008) for the viscosity of the ice shell ($10^{13}-10^{14}$ 
Pa s). It is also close to the reference viscosity of water at 255 K ($10^{15}$ Pa s) adopted by B\u{e}hounkov\'a et al. (2012) in their modeling of the 
melting events at origin of the south-pole activity on Enceladus. More recent research carried out by \v{C}adek et al. (2019) demonstrates that the 
viscosity of ice at the melting temperature may be equal to or higher than $3\times10^{14}\ {\rm Pa\ s}$, for the ice shell to remain stable.

The forced libration obtained with the creep tide theory for homogeneous bodies, in the case of Enceladus, is 3.1 times smaller than the libration amplitude 
obtained from Cassini's observations. However, results close to the observation were obtained in a preliminary extension of the core-shell model developed by 
the authors (Folonier, 2016; Folonier and Ferraz-Mello, 2017; Folonier et al., in preparation), in which one liquid layer is assumed to exist between the 
crust and the core.

In the found range of values of $\gamma$, the contribution of the satellite tides to the variations of the semi-major axis and eccentricity of Enceladus are 
$\langle\dot{a}\rangle=-(0.4-1.3)\times 10^{-5}\ {\rm km/yr}$ and $\langle \dot{e}\rangle=-(1.9-6.0)\times 10^{-9}\ {\rm yr}^{-1}$. The eccentricity 
variation is very small. Nonetheless, in the case of Enceladus, we need to consider the effects of the almost 2:1 resonance between Enceladus and 
Dione, that produces a forced eccentricity of 0.00459 (see Ferraz-Mello, 1985; Vienne and Duriez, 1995). In the case of Mimas, we have found that the 
variations of the semi-major axis and eccentricity are $\langle\dot{a}\rangle\sim - 10^{-6}\ {\rm km/yr}$ and $\langle \dot{e}\rangle\sim- 10^{-10}\ {\rm yr}^{-1}$. 
In both cases, the variations found are much smaller than those due to the tides on the planet.

This paper is organized as follows: We first proceed, in Sections 2 to 4, to a revisit of the creep tide theory proposing a new approach in which the 
differential equations for the tidal deformation of the primary, and for its rotation, are integrated simultaneously. Then, in Section 5, we study the case 
in which the rotation of the primary is nearly synchronous but not uniform showing a forced libration. In Section 6 and 7, we do an inventory of the main 
mechanical processes involving the storage of mechanical energy in the system and discuss the law ruling the dissipation in stiff satellites, with 
application to Enceladus. In Section 8,  we extend the results to obtain the average variations of the metrical elements of the orbit: semi-major axis and 
eccentricity. In the final sections, we extend the results to Mimas and present the conclusions. The paper is completed by two appendices, where are given 
some technical details of the model used to evaluate the orbital energy (Appendix 1) and the analytical approximation of the spin-orbit quasi-synchronous 
attractor (Appendix 2). In addition, an Online Supplement is provided, where the classical creep tide approach is also extended to include the 
near-synchronous, but not uniform, rotation case.

\section{The creep tide equations}\label{sec02}

Let us consider one system formed by the extended body $\tens{m}$ (primary) and the mass point $\tens{M}$ (companion) and let $\textbf{r}$ be the 
radius-vector in a system of reference centered on \tens{m}. We assume that the primary is a homogeneous body and have an angular velocity of rotation 
$\vec{\Omega}$, perpendicular to the orbital plane.

\begin{figure}[h]
\begin{center}
 \includegraphics[scale=0.35]{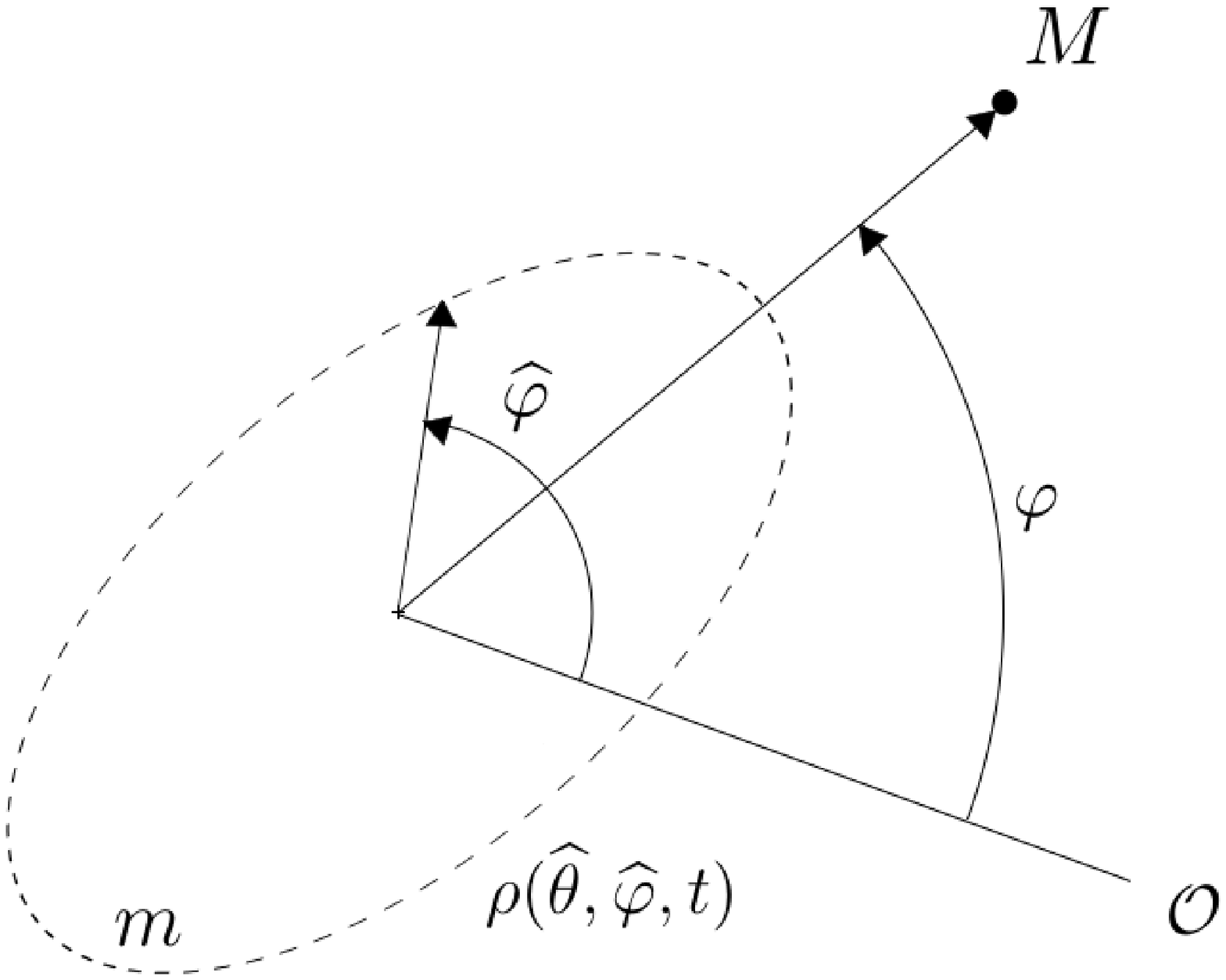}\ \ \ \ \ \ \ \ \ \ \ 
 \includegraphics[scale=0.35]{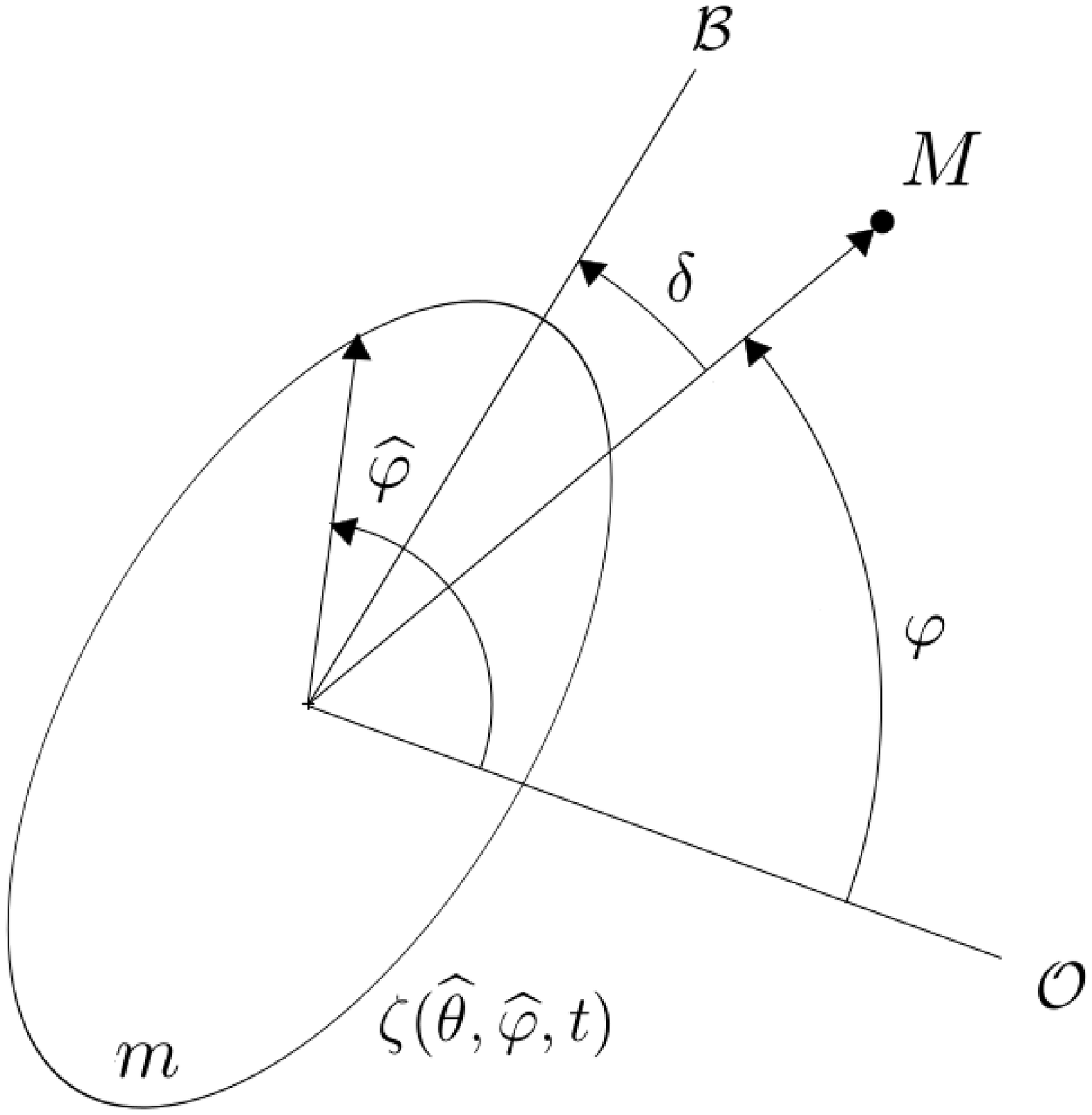}
\caption{Equatorial sections of the ellipsoids $\rho(\widehat{\theta},\widehat{\varphi},t)$ and $\zeta(\widehat{\theta},\widehat{\varphi},t)$ corresponding 
to the tide on \tens{m} generated by \tens{M}. Angles: $\varphi$ and $\widehat\varphi$ are the longitudes, in a fixed reference system, of the companion and 
of one point on the surface of the primary, respectively. $\delta$ is the orientation angle of the tidal bulge (labeled $\mathcal{B}$) with respect to the 
companion.}
\label{fig01}
\end{center}
\end{figure}

In the creep tide theory, the tidal deformation of \tens{m} is obtained by solving the Newtonian creep law
\begin{equation}
\dot{\zeta} = \gamma (\rho-\zeta),
\label{eq:ansatz} 
\end{equation}
where $\gamma$ is the relaxation factor (see paper I), $\zeta=\zeta(\widehat\varphi,\widehat\theta,t)$ is the distance of the surface point of coordinates $\widehat\varphi$ (longitude) and 
$\widehat\theta$ (co-latitude) to the center of gravity of the body, and  $\rho=\rho(\widehat\varphi,\widehat\theta,t)$ is the surface of the 
static figure of equilibrium of $\tens{m}$ under the gravitational attraction of \tens{M}. $\rho$ is approximated by a triaxial ellipsoid whose major axis
is oriented towards  $\tens{M}$, and whose equatorial prolateness and polar oblateness are 
\begin{equation}
\epsilon_\rho=\frac{a-b}{R_e}=\frac{15MR_e^3}{4mr^3},
\label{eq:prola}
\end{equation}
and 
\begin{equation}
\epsilon_z=1-\frac{c}{R_e}=\frac{\epsilon_\rho}{2}+\frac{5\Omega^2R_e^3}{4Gm},
\label{eq:obla}
\end{equation}
where $a,b,c$ are the semi-major axes of the triaxial ellipsoid, $R_e$ is the mean equatorial radius, $G$ is the gravitational constant and $\Omega$ is the 
spin rate of \tens{m} (see Tisserand, 1891; Chandrasekhar 1969; Folonier et al. 2015). Its equation is\footnote{In paper II (Ferraz-Mello, 2015), the 
variation of $R_e$ has been neglected. However, when the equatorial prolateness varies due to a variation in the distance of \tens{m} to \tens{M}, the polar 
flattening and $R_e$ vary accordingly: $R_e \simeq R(1+\frac{1}{3}\epsilon_z)$ where $R$ is the mean radius of the primary (constant).}
\begin{equation}
\rho(\widehat{\theta},\widehat{\varphi},t) = R \left(1+\frac{1}{2}\epsilon_\rho \sin^2{\widehat{\theta}}\cos{(2\widehat{\varphi}-2\varphi)}+\epsilon_z\bigg(\frac{1}{3}-\cos^2{\widehat{\theta}}\bigg)\right),
\label{eq:rho_eq}
\end{equation}
where $\varphi$ is the true longitude of the companion in its equatorial orbit around the primary and $R$ is the mean radius of the primary. The angles are 
such that the major axis of the ellipsoid is always oriented towards the companion (\textit{left} panel of Fig. \ref{fig01}). The right-hand side is a time 
function depending on the longitude $\widehat\varphi$ (such that $d\widehat\varphi/dt = \Omega(t)$) and on the polar coordinates of the companion, $r$ and 
$\varphi$. The radius vector of \tens{M}, $r$, is introduced in the equation by the flattenings $\epsilon_\rho$ and $\epsilon_z$.

In previous papers (papers I and II) the forced terms in the solution of Eq. (\ref{eq:ansatz}) were approximated by the sum of an arbitrary number of 
ellipsoidal bulges over one sphere of radius $R$ (see papers I and II), where each bulge has different flattenings and orientation. To the first order in the 
flattenings, the sum of two or more ellipsoidal bulges can be expressed by a new ellipsoidal bulge with its flattenings and orientation (see Appendix 3 of 
Folonier and Ferraz-Mello, 2017). For this reason, $\zeta$ can be approximated by a homogeneous ellipsoid:
\begin{eqnarray}
\zeta(\widehat{\theta},\widehat{\varphi},t) &=& R\left(1+\frac{1}{2}\mathcal{E}_\rho \sin^2{\widehat{\theta}}\cos{(2\widehat{\varphi}-2\varphi_\mathcal{B})}+\mathcal{E}_z\bigg(\frac{1}{3}-\cos^2{\widehat{\theta}}\bigg)\right),
\label{eq:zeta-pro}
\end{eqnarray}
where the instantaneous flattenings $\mathcal{E}_\rho$, $\mathcal{E}_z$ and the orientation longitude of the bulge $\varphi_\mathcal{B}=\varphi+\delta$ are 
unknown functions of the time. Here, $\delta$ is the orientation angle of the bulge vertex with respect to \tens{M} (\textit{right} panel of Fig. \ref{fig01}).

In order to find the time evolution of these unknown functions, we differentiate (\ref{eq:zeta-pro}) with respect to the time and replace the result into 
the creep equation (\ref{eq:ansatz}). We obtain:
\begin{eqnarray}
&&\left(\Big(\dot{\mathcal{E}}_\rho + \gamma\mathcal{E}_\rho\Big)\cos{2\delta}+\mathcal{E}_\rho(2\Omega-2\dot\varphi-2\dot{\delta})\sin{2\delta}\right)\frac{1}{2} R \sin^2{\widehat{\theta}}\cos{(2\widehat{\varphi}-2\varphi)}\nonumber\\
&&+\left(-\mathcal{E}_\rho(2\Omega-2\dot\varphi-2\dot{\delta})\cos{2\delta}+\Big(\dot{\mathcal{E}}_\rho + \gamma\mathcal{E}_\rho\Big)\sin{2\delta}\right)\frac{1}{2} R \sin^2{\widehat{\theta}}\sin{(2\widehat{\varphi}-2\varphi)}\nonumber\\
&&+R\left(\dot{\mathcal{E}}_z+\gamma\mathcal{E}_z\right)\left(\frac{1}{3}-\cos^2{\widehat{\theta}}\right)= \frac{1}{2} R \sin^2{\widehat{\theta}}\gamma\epsilon_\rho\cos{(2\widehat{\varphi}-2\varphi)}+R\gamma\epsilon_z\left(\frac{1}{3}-\cos^2{\widehat{\theta}}\right).
\label{eq:creep}
\end{eqnarray}

Since the flattenings $\mathcal{E}_\rho,\mathcal{E}_z$ and the angle of orientation $\delta$ cannot depend on one specific point on the surface of \tens{m} 
(that is $\widehat{\theta},\widehat{\varphi}$), we can decompose Eq. (\ref{eq:creep}) into three equations, one for each trigonometric argument, which may 
be written as
\begin{eqnarray}
 \dot{\delta}          &=& \Omega-\dot{\varphi}-\frac{\gamma\epsilon_\rho}{2\mathcal{E}_\rho} \sin{2\delta} \nonumber\\
\dot{\mathcal{E}}_\rho &=& \gamma\Big(\epsilon_\rho \cos{2\delta}-\mathcal{E}_\rho\Big) \nonumber\\
 \dot{\mathcal{E}}_z   &=& \gamma\Big(\epsilon_z-\mathcal{E}_z\Big).
\label{eq:system-creep}
\end{eqnarray}
This system of differential equations of first order allows us to calculate the time evolution of the instantaneous flattenings and the instantaneous 
orientation angle, when the orbital motion of the companion (that is, $r$ and $\dot{\varphi}$) and the spin rate $\Omega$ are known. This new version of the 
creep equations is formally analogous to the one used by Correia et al. (2014), Bou\'{e} et al. (2016), Correia et al. (2018) and Beaug\'{e} (personal 
communication).

\section{The disturbing potential}

The disturbing potential created by the homogeneous triaxial ellipsoid \tens{m}, the bulge of which is rotated of an angle $\varphi_\mathcal{B}$ with 
respect to the axis $x$, at the companion \tens{M}$(r,\theta=\pi/2,\varphi)$, neglecting the harmonics of degree higher than 2, is:
\begin{eqnarray}
 \delta U(\vec{r}) &=& -\frac{G(B-A)}{2r^3}\Big(3\cos^2{\Psi_\mathcal{B}}-1\Big)-\frac{G(C-B)}{2r^3},
\end{eqnarray}
where $G$ is the gravitational constant, $\Psi_\mathcal{B}$ is the angle between the direction of the point where the potential is taken and the direction 
of the bulge vertex, labeled by $\mathcal{B}$ (see Fig. \ref{fig01}), and $A,B,C$ are the moments of inertia of the ellipsoid with respect to its principal 
axes ($A<B<C$).

The cosine can be written as 
\begin{eqnarray}
\cos{\Psi_\mathcal{B}} = \vec{\widehat{r}}\cdot\vec{\widehat{r}}_\mathcal{B},
\end{eqnarray}
where $\vec{\widehat{r}}$ and $\vec{\widehat{r}}_\mathcal{B}$ are the unitary vectors oriented towards the companion and the bulge direction, respectively. 
The differences $B-A$ and $C-B$, to the first order in the flattenings, can be approximated by
\begin{equation}
 B-A\approx C\mathcal{E}_\rho; \ \ \ \ \ \ \ C-B\approx C\left(\mathcal{E}_z-\frac{1}{2}\mathcal{E}_\rho\right).
\end{equation}
The moment of inertia with respect to the polar axis is $C=m(a^2+b^2)/5$ or, introducing the flattenings of the ellipsoid resulting from the integration of 
the creep differential equation
\begin{equation}
C=\frac{2}{5}mR^2\left(1+\frac{2}{3}\mathcal{E}_z  + {\cal O}({\mathcal{E}^2})\right).
\end{equation}

Hence, using the definition of $\epsilon_\rho$, the resulting disturbing potential is
\begin{eqnarray}
 \delta U(\vec{r}) &=& -\frac{2Gm^2\epsilon_\rho}{25MR}\mathcal{E}_\rho\bigg(2(\vec{\widehat{r}}\cdot\vec{\widehat{r}}_\mathcal{B})^2-1\bigg)-\frac{4Gm^2\epsilon_\rho}{75MR}\mathcal{E}_z.
\end{eqnarray}

It is important to emphasize that, since the beginning of this paper, we have used the same notation $\mathbf{r}$ to indicate the radius vector of the 
companion in all situations. We have thus broken with the tradition of using $\mathbf{r}^*$ (or $\mathbf{r}'$) to indicate the radius vector of the 
companion in the equations used to calculate the tidal deformation of the primary (Darwin, 1880; Kaula, 1964; MacDonald, 1964; Efroimsky, 2012). The 
dichotomy $\mathbf{r}, \mathbf{r}^*$ was introduced by Darwin for the only reason that, in the calculus of the force as the gradient of the potential $U$, 
the derivatives of the potential  must be done with respect to the coordinates of the mass point where the force is applied. Thus, it was important to use a 
different notation for the radius vector and its  components in the calculation of the flattenings, and to write the potential as a function $U(\mathbf{r}, 
\mathbf{r}^*)$ making explicit which of the radii vectors was to be considered when the force is calculated. But, physically, there is only one radius 
vector being considered and, after the gradient calculation, $\mathbf{r}$ and $\mathbf{r}^*$ are identified. As in Correia et al. (2014), Bou\'{e} et al. 
(2016), Ragazzo and Ruiz (2017) and Correia et al. (2018), in the approach proposed in this paper, the potential does not depend explicitly on both 
instances of the radius vector. The radius vector used in the calculation of the flattenings is substituted by the time functions ${\cal E}_\rho$ and 
${\cal E}_z$ and the only radius vector $\mathbf{r}$ appearing in Eq. (\ref{eq:system-creep}) is the radius vector of the point where the force is being 
applied.

\section{The tidal force and torque}\label{sec04}

To calculate the tidal force
$\vec{F}$ acting on the mass $M$ located in \tens{M}, we take the negative gradient of the disturbing potential of $\tens{m}$ and
multiply it by the mass placed in the point. Hence
\begin{equation}
\vec{F}=-M\nabla_{\vec{r}} \delta U. 
\end{equation}
The sign in this expression comes from the fact that we are using the conventions of Physics ($\delta U$ is a potential not a force-function). It is important to 
stress that in agreement with Newton laws, there exists a reaction force $-\vec{F}$ acting on \tens{m} due to the attraction of \tens{M} (see Ferraz-Mello 
et al. 2003). This fact is generally neglected in studies where one of the masses is negligible when compared to the other but its neglect in general 
problems is an error. Then, we obtain:
\begin{equation}
 \vec{F} =-\frac{2Gm^2\epsilon_\rho}{25Rr^2}\mathcal{E}_\rho\bigg(10(\vec{\widehat{r}}\cdot\vec{\widehat{r}}_\mathcal{B})^2\vec{r}-4r(\vec{\widehat{r}}\cdot\vec{\widehat{r}}_\mathcal{B})\vec{\widehat{r}}_\mathcal{B}-3\vec{r}\bigg)-\frac{4Gm^2\epsilon_\rho}{25Rr^2}\mathcal{E}_z\vec{r}.
\end{equation}

The unitary vector $\vec{\widehat{r}}_\mathcal{B}$ can be decomposed in terms of the unitary vectors $\vec{\widehat{r}}$ and $(\vec{\widehat{z}}\times\vec{\widehat{r}})$, 
where $\vec{\widehat{r}}$ and $\vec{\widehat{z}}$ are the unitary vectors oriented towards \tens{M} and along the $z$-axis, respectively: 
\begin{equation}
\vec{\widehat{r}}_\mathcal{B} =  \cos{\delta}\ \vec{\widehat{r}}+ \sin{\delta}\ (\vec{\widehat{z}}\times\vec{\widehat{r}}).
\end{equation}

Hence, the resulting tidal force acting on \tens{M} can be written as:
\begin{equation}
 \vec{F} = -\frac{2Gm^2\epsilon_\rho}{25Rr^2}\mathcal{E}_\rho\bigg(3\cos{2\delta}\ \vec{r}-2\sin{2\delta}\ (\vec{\widehat{z}}\times\vec{r})\bigg)-\frac{4Gm^2\epsilon_\rho}{25Rr^2}\mathcal{E}_z\vec{r}.
\label{eq:force}
\end{equation}

Finally, the tidal torque acting on \tens{M} is $\vec{M}=\vec{r}\times\vec{F}$, or:
\begin{equation}
\vec{M}= \frac{4Gm^2\epsilon_\rho}{25R}\mathcal{E}_\rho\sin{2\delta}\ \vec{\hat{z}}.
\label{eq:torque}
\end{equation}
It is important to note that, in order to calculate the angular acceleration of the primary, we need to consider the reaction on the primary, that is 
\begin{equation}
-\vec{M}=C\dot{\vec{\Omega}}+\dot{C}\vec{\Omega},
\end{equation}
where the time variation of the axial moment of inertia is
\begin{equation}
\dot{C}\approx\frac{4}{15}mR^2 \dot\mathcal{E}_z.
\end{equation}
We note that, at this order of approximation, the equatorial tidal prolateness does not affect the moment of inertia (the deformations inward and outward 
compensate themselves).

The scheme considered in this approach differs from the scheme used in papers I and II. Here, the equations of the instantaneous ellipsoidal bulge and its 
orientation may be integrated together with the rotational equation, while, in papers I and II, the shape of the primary is calculated assuming $\Omega$ 
as a known time function through the longitude $\widehat{\varphi}=\Omega(t-t_0)$. Once the shape has been determined, it is used to obtain the rotational 
evolution of the deformed body.

The model presented here allows us also to calculate the orbital evolution of the companion, as well as the evolution of the ellipsoidal bulge, its 
orientation and the rotational evolution of the primary. However, in general, the variation of the orbital elements is much slower and, for short time spans, 
we may assume a Keplerian motion for the companion. The instantaneous shape, orientation and rotation of the primary are then calculated using Eqs. 
(\ref{eq:system-creep}) and (\ref{eq:torque}) and the classical two-body expressions for $r$ and $\dot{\varphi}$.

\section{Solution in the neighborhood of the synchronous rotation.}\label{sec05}

In the neighborhood of the synchronization, the rotation of close-in satellites and exoplanets is generally damped towards a final stable state which 
depends on the nature of the body. The rotation of gaseous planets, of fast relaxation (high $\gamma$, low viscosity), tends to a stationary rotation 
slightly faster than the orbital motion (a.k.a. supersynchronous motion). The excess of angular velocity is $\sim 6ne^2$ (see paper I). On the other end, 
the rotation of planetary satellites and Earth-like planets, of slow relaxation (low $\gamma$, high viscosity), is damped to attractors with the same period 
as the orbital period and the final rotations are not uniform. They are forced oscillations (physical librations) around one center. In this case, the 
solution of the creep equation can no longer be calculated as in papers I and II. The use of the uniform approximation for $\Omega$ is no longer appropriate 
because the rotation is, in this case, affected by a significant short-period oscillation\footnote{The extension of the theory of papers I and II to the 
case in which the rotation is trapped in a periodic attractor is given in the Online Supplement linked to this paper.}. It is important to emphasize that in 
the approach adopted in this paper, no hypotheses on the rotation behavior are necessary since all equations are integrated simultaneously. The synchronous 
attractor may be approximated by
\begin{equation}
\nu \defeq 2(\Omega-n) \cong B_0 + B_1\cos{\ell}+ B_2\sin{\ell},
\end{equation}
where $\ell$ is the mean anomaly of the companion; the mean value $B_0$ and the amplitudes $B_1,B_2$ are constants (for the details of the calculation of 
the synchronous attractor, see Appendix 2).

\begin{figure}[h]
\begin{center}
 \includegraphics[height=55mm,clip=]{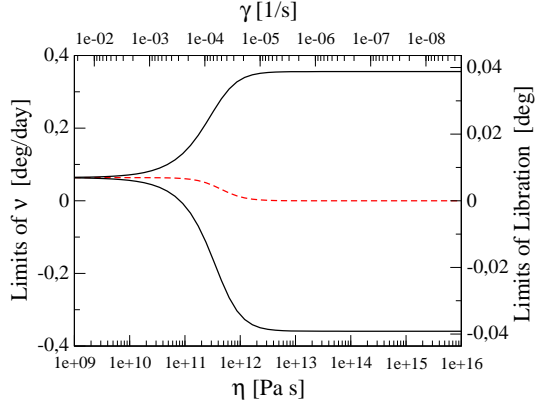}
\caption{Tidal forced oscillation of the semi-diurnal frequency of one body like Enceladus in function of the viscosity $\eta$ and the relaxation factor 
$\gamma$ (shown on the upper axis). The black lines are the limits of the oscillation. The red dashed line is the mean value of $\nu$. In the right axis we 
show the corresponding limit of the physical libration when a harmonic oscillation is assumed.}
\label{fig02}
\end{center}
\end{figure}

In the case of Enceladus, the numerical solution of the exact equations for low values of $\gamma$ is an almost symmetric oscillation of $\nu$. The 
semi-amplitude of this oscillation depends on the adopted relaxation factor (see Fig. \ref{fig02}). If $\gamma< 10^{-5}\ {\rm s}^{-1}$, the forced 
oscillation amplitude is $0.36\ {\rm deg/d}$.

This value may be compared to the observed values (see Table \ref{tab:resu}). Measurements of control points on the surface of Enceladus accumulated over 
seven years of Cassini's observations allowed Thomas et al. (2016) to determine the satellite's rotation state. They have found a libration of 0.120 $\pm$ 
0.014 degrees. If we assume that these oscillations follow a harmonic law we obtain, correspondingly, for the oscillation of the velocity of rotation: $0.56 
\pm 0.06\ {\rm  deg/day}$, and for the semi-diurnal frequency $1.12 \pm 0.12\ {\rm  deg/day}$. The immediate conclusion from the comparison
of these values is that it is not possible to reproduce exactly the observed forced libration of Enceladus with a homogeneous body model. The predicted 
oscillation is smaller. However, results close to the observation were obtained in a preliminary extension of the core-shell model developed by the authors, 
when one liquid layer is assumed to exist between the crust and the core (Folonier, 2016; Folonier and Ferraz-Mello, 2017; Folonier et al., in preparation).

For the sake of comparison, we may note that viscoelastic models adopting a permanent triaxiality give a result yet smaller: $\lesssim 0.03\ {\rm deg}$ 
depending on the adopted triaxiality (Rambaux et al., 2010).
\begin{table}[t]
\caption{Satellites data used in the calculations}
\begin{tabular}{lcc}
\hline 
 & {Enceladus} & {Mimas } \\
\hline
\\ 
Mass ($10^{20}$ kg) & 1.08 &0.379 \\
Mean Radius (km)&252.1&198.2\\
Sidereal Period (d)& 1.370218 & 0.942422\\ 
Semi-major axis ($10^3$ km)&238.02 & 185.52 \\
Eccentricity & 0.0045 & 0.01986 \\
Mean-motion ($10^{-5}{\rm s}^{-1}$)&5.300508&7.696292\\
\hline
\label{tab:data}
\end{tabular}
\end{table}

\section{The energy balance}\label{sec06}

In this section, we examine the various manifestations of the mechanical energy in a system formed by two mutually attracting bodies: the extended body 
$\tens{m}$ and one mass point $\tens{M}$. Let their masses be respectively, $m$ and $M$. We consider only the case where $\tens{M}$ lies on the equatorial 
plane of $\tens{m}$. Our aim is to evaluate the amount of the energy dissipated by the body\footnote{The dissipation in each one of the two bodies may be 
considered separately. Within the order of approximation generally adopted (first order in the tidal deformations), the variation of the energy can be split 
into two parts, each one associated with the tidal deformation in one of the bodies while the other - source of the tidal potential - is kept as a mass 
point. Therefore, only the dissipation in one of the two bodies need to be explicitly considered.}.

Let us first review some known facts of a system formed by the extended body $\tens{m}$ and the mass point $\tens{M}$ (see Scheeres, 2002). Let $\textbf{r}, 
\textbf{V}$ be the radius-vector and the velocity of \tens{M} in a system of reference centered on \tens{m}. The kinetic energy referred to the center of 
gravity of the system formed by the two bodies and the rotational energy of $\tens{M}$ are
\begin{equation}
E_{\rm kin}=\half\frac{Mm}{M+m}\mathbf{V}^2; \ \ \ \ \ \ E_{\rm rot} = \frac{1}{2}C\Omega^2,
\end{equation}
the time-dependent gravitational potential generated by the primary in the point $\mathbf{r}$ is
\begin{eqnarray}
MU\big(\mathbf{r};t\big) &=& -\frac{GMm}{r}-\frac{2Gm^2}{25R}\left(\epsilon_\rho\mathcal{E}_\rho\cos{\big(2\varphi-2\varphi_\mathcal{B}\big)}-\frac{2}{3}\epsilon_\rho\mathcal{E}_z\right),
\end{eqnarray}
and $E_{\rm int}$ is the internal gravitational energy of the primary. Using the explicit formulas given by Ess\'en (2004) (see Appendix 1), we obtain
\begin{eqnarray}
E_{\rm{int}} &=& -\frac{3Gm^2}{5R}\left(1-\frac{1}{15}\mathcal{E}_\rho^2-\frac{4}{45}\mathcal{E}_z^2\right).
\end{eqnarray}

The time derivatives of the energies are
\begin{equation}
 \dot{E}_{\rm kin}= \frac{Mm}{M+m}\mathbf{V}\cdot \dot{\mathbf{V}}=\mathbf{f}\cdot\mathbf{V}; \ \ \ \ \ \ \dot{E}_{\rm rot} = -\vec{M}\cdot\vec{\Omega}-\frac{1}{2}\dot{C}\Omega^2
 \label{eq:dot E-kin}
\end{equation}
(since in Newton's law the acceleration must be referred to the barycenter, that is, $\mathbf{f} = \frac{Mm}{M+m} \dot{\mathbf{V}}$)
\begin{equation}
M\dot{U}= M{\rm grad}_\mathbf{r}U \cdot \mathbf{V} + M\frac{\partial U}{\partial t} = - \mathbf{f}\cdot\mathbf{V} + M\frac{\partial \delta U}{\partial \varphi_\mathcal{B}} \dot\varphi_\mathcal{B} +M\frac{\partial \delta U}{\partial \mathcal{E}_\rho} \dot\mathcal{E}_\rho+M\frac{\partial \delta U}{\partial \mathcal{E}_z} \dot\mathcal{E}_z,
 \label{eq:dot E-pot}
\end{equation}
and
\begin{equation}
\dot{E}_{\rm int} = \frac{\partial E_{\rm{int}}}{\partial \mathcal{E}_\rho} \dot\mathcal{E}_\rho+\frac{\partial E_{\rm{int}}}{\partial \mathcal{E}_z} \dot\mathcal{E}_z.
 \label{eq:dot E-int}
\end{equation}

Hence,
\begin{eqnarray}
\dot{E}_{\rm tot} &=& \dot{E}_{\rm kin} + M\dot{U} + \dot{E}_{\rm int} + \dot{E}_{\rm rot},
\end{eqnarray}
or
\begin{eqnarray}
\dot{E}_{\rm tot} &=& M\frac{\partial \delta U}{\partial \varphi_\mathcal{B}} \dot\varphi_\mathcal{B} +M\frac{\partial \delta U}{\partial \mathcal{E}_\rho} \dot\mathcal{E}_\rho+M\frac{\partial \delta U}{\partial \mathcal{E}_z} \dot\mathcal{E}_z + \frac{\partial E_{\rm{int}}}{\partial \mathcal{E}_\rho} \dot\mathcal{E}_\rho+\frac{\partial E_{\rm{int}}}{\partial \mathcal{E}_z} \dot\mathcal{E}_z+  \dot{E}_{\rm rot},
\label{eq:dotE}
\end{eqnarray}
that is, the energy variation associated with the power $\mathbf{f}\cdot\mathbf{V}$ is exchanged with the kinetic energy and cannot account for the 
dissipation of the system energy (see Eqs. (\ref{eq:dot E-kin}) and (\ref{eq:dot E-pot})).
 This is a well-known fact in the study of the motion of satellites around non-spherical rigid bodies\footnote{It is worth 
stressing that we are referring to the actual energy of the system. At variance with it, the energy of the osculating Keplerian motion may vary (as well as 
the osculating semi-major axis) but such variation is only a consequence of the way in which osculating variables are defined.}.

Using Eqs. (\ref{eq:system-creep}), the time derivative of the total mechanical energy can be written as
\begin{eqnarray}
\dot{E}_{\rm tot} &=& -\frac{6}{5}\frac{Gm^2}{\gamma R}\left(\frac{1}{15}\dot{\mathcal{E}}_\rho^2+\frac{1}{15}\mathcal{E}_\rho^2\big(2\Omega-2\dot{\varphi}-2\dot{\delta}\big)^2+\frac{4}{45}\dot{\mathcal{E}}_z^2\right)\le0.
\label{eq:dotE_tot}
\end{eqnarray}
The time derivative of the total mechanical energy is always negative.

Finally, the orbital energy is:
\begin{equation}
E_{\rm orb} = E_{\rm kin} + MU\big(\mathbf{r};t\big),
\label{eq:E_orb}
\end{equation}
and, using Eqs. (\ref{eq:dot E-kin})-(\ref{eq:dot E-pot}), the time derivative of the orbital energy is
\begin{equation}
\dot E_{\rm orb} = \dot E_{\rm kin} + M\dot U\big(\mathbf{r};t\big),
\label{eq:dotE_orb}
\end{equation}
This definition of the orbital energy variation is the same as adopted by Correia et al (2014).

\section{Dissipation}\label{sec07}

\begin{figure}[h]
\begin{center}
 \includegraphics[height=55mm,clip=]{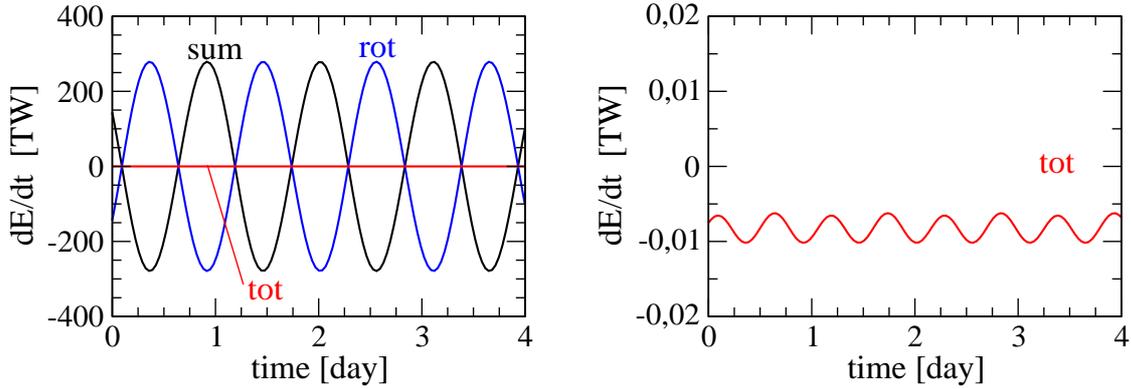}
\caption{{\it Left:} Tidal variation of the mechanical energy (after the vanishing of the transients): the sum of the orbital energy and the internal 
gravitational energy (black line labeled \textit{sum}), the rotational (blue line labeled \textit{rot}), and the total mechanical energy (red line labeled \textit{tot}). {\it 
Right:} $\dot{E}_{\rm tot}$ in a magnified scale. N.B. $\gamma= 2 \times 10^{-7} {\rm s}^{-1}$.}
\label{fig03}
\end{center}
\end{figure}

The actual variation of the main components of the mechanical energy is shown in the left panel of Fig. \ref{fig03}.  It is large, indicating a great 
periodic energy exchange between the rotational energy (labeled \textit{rot}) and the sum of the orbital energy and the internal gravitational energy (labeled \textit{sum}). 
The amplitudes of variation of the orbital and rotational energies have the same order of magnitude and the variation of the total energy (labeled 
\textit{tot} in Fig. \ref{fig03}), shows only a very small variation (of the order of one part in $10^5$ of the variation of the two components). This very 
small periodic variation of the time derivative of the total mechanical energy is always negative (right panel of Fig. \ref{fig03}).

In the long run, the only sources for the energy dissipated by the tides are the orbital and the rotational energies. The usual operation to get rid of 
periodic variations is the averaging of the total mechanical energy. The canonical tool to separate conservative and dissipative terms is the analysis of 
the differential form expressing the variation of the orbital energy. However, in the adopted model, only the attraction of the external body $\tens M$ by 
the deformed body $\tens m$ is available. The dynamics of the action of $\tens M$ creating the deformation in $\tens m$ is concealed by the creep equation 
used to determine the shape of $\tens m$. The averaging to zero of the periodic variations may be considered as the equivalent of the zero variation of the 
energy on a closed path characteristic of the conservative phenomena.

\subsection{Analytical approximation}

\begin{figure}[h]
\begin{center}
 \includegraphics[height=55mm,clip=]{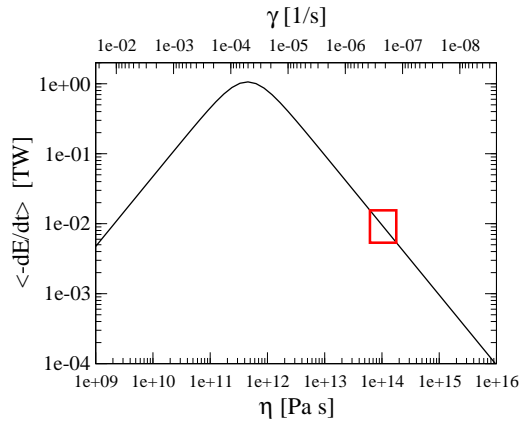}
\caption{Dissipation curve: Average of the net variation of the mechanical energy. The body and orbital parameters used correspond to a homogeneous 
Enceladus. The actual range of the observed dissipation and the corresponding ranges for the viscosity (and for the relaxation factor) are shown by a 
red box.}
\label{fig04}
\end{center}
\end{figure}

The variable characterizing the variation in a short time interval is the mean anomaly $\ell$ and the averaging operation is just $\frac{1}{2\pi}\int_0^{2\pi} 
\dot{E}_{\rm tot} \D\ell$. In the case of Enceladus, the result is shown in Fig. \ref{fig04}. We note that the resulting $\langle\dot{E}_{\rm tot}\rangle = 
\langle \dot{E}_{\rm orb}+\dot{E}_{\rm rot}+\dot{E}_{\rm int} \rangle$ (black solid line) is negative, hence, the system is losing mechanical energy, as 
expected. Using the analytical solution detailed in Appendix 2, the time average of the total mechanical energy (given by Eq. \ref{eq:dotE_tot}), for the 
synchronous attractor, can be approximated as
\begin{equation}
\langle\dot{E}_{\rm tot}\rangle_{sync} = -\frac{21GMmR^2\overline{\epsilon}_\rho e^2}{5a^3}\frac{n^2\gamma}{n^2+\gamma^2},
\label{eq:<dotE_tot>}
\end{equation}
to the first order in the flattenings.

For sake of completeness, it is worth repeating some properties of this result: (1) the result is always negative (energy is lost); (2) the variation of the 
dissipation with the relaxation factor has the inverted V-shape, characteristic of the Maxwell rheology; (3) in the neighborhood of the stationary solution, 
the quantity defined by Eq. (\ref{eq:<dotE_tot>}) is of the order ${\cal O}(e^2)$.

In the case of Enceladus, the estimations of the heat dissipated in the SPT (south polar terrain) area based on the observations with Cassini are in the 
range $5-16$ GW (\textit{cf} Howett et al. 2011; Spencer et al. 2013; Le Gall et al. 2017). This observed dissipation corresponds to a relaxation factor 
$\gamma = 1.2-3.8 \times 10^{-7}$ s$^{-1}$ (red box in Fig. \ref{fig04}). The viscosity corresponding to this relaxation\footnote{According with the 
relation $\gamma=wR/2\eta$ given in papers I and II ($w$ is the specific weight at the surface of the body).} is $0.6-1.9 \times 10^{14}$ Pa s. The physical libration is responsible for a 27 percent increase in 
the dissipation of Enceladus (See Section 5 in the Online Supplement). This last result is in good agreement with the value 30 percent found by Efroimsky 
(2018).

\section{Semi-major axis and eccentricity average perturbations}

\subsection{Semi-major axis}

The variation of the osculating semi-major axis due to the tides raised on $\tens{m}$ may be obtained using the corresponding Lagrange variational equation:
\begin{equation}
\dot{a}= \frac{2}{na} \frac {\partial {\cal R}}{\partial \ell},
\end{equation}
where the disturbing function is ${\cal R}=-(1+M/m)\delta U$ (see Brouwer and Clemence, 1961, Chap. XI).  The minus sign is included because $\delta U$ is a 
potential (not a force-function) and the factor $(1+M/m)$ is introduced to account for the fact that the disturbing force is not of external origin but an 
interaction between the two bodies. We thus consider the force per unit mass acting on one body minus its reaction on the other body (see discussion in 
Ferraz-Mello et al. 2008, Section 18).

Hence, considering the third Kepler law ($n^2a^3=G(M+m)$) and comparing to the work calculated above,
\begin{equation}
\dot{a}=\frac{2a^2}{GmM}\dot E_{\rm orb}.
\label{eq:dadt1}
\end{equation}
This is the same equation obtained when taking the time derivatives of both sides of the two-body classical equation relating the orbital energy and the 
osculating semi-major axis,  $E_{\rm orb}= -GmM/2a$ (see Brouwer and Clemence, 1961), which has been used in previous papers (see paper I). It 
shows that $\dot E_{\rm orb}$ is equal to the time derivative of the Keplerian orbital energy of the system.

\begin{figure}[h]
\begin{center}
 \includegraphics[height=55mm,clip=]{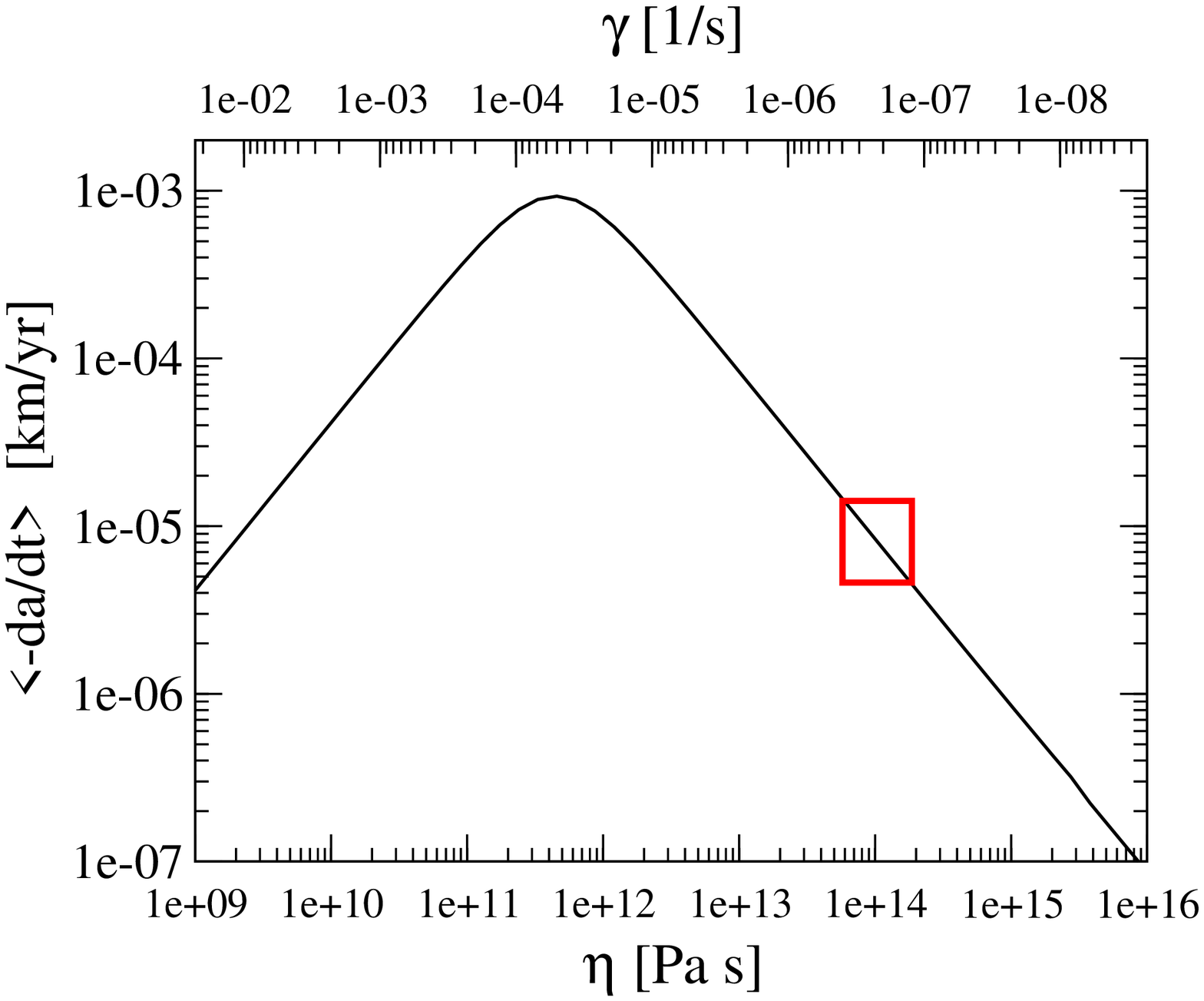}\ \ \ \ \ \ \ \ \ \ \ 
 \includegraphics[height=55mm,clip=]{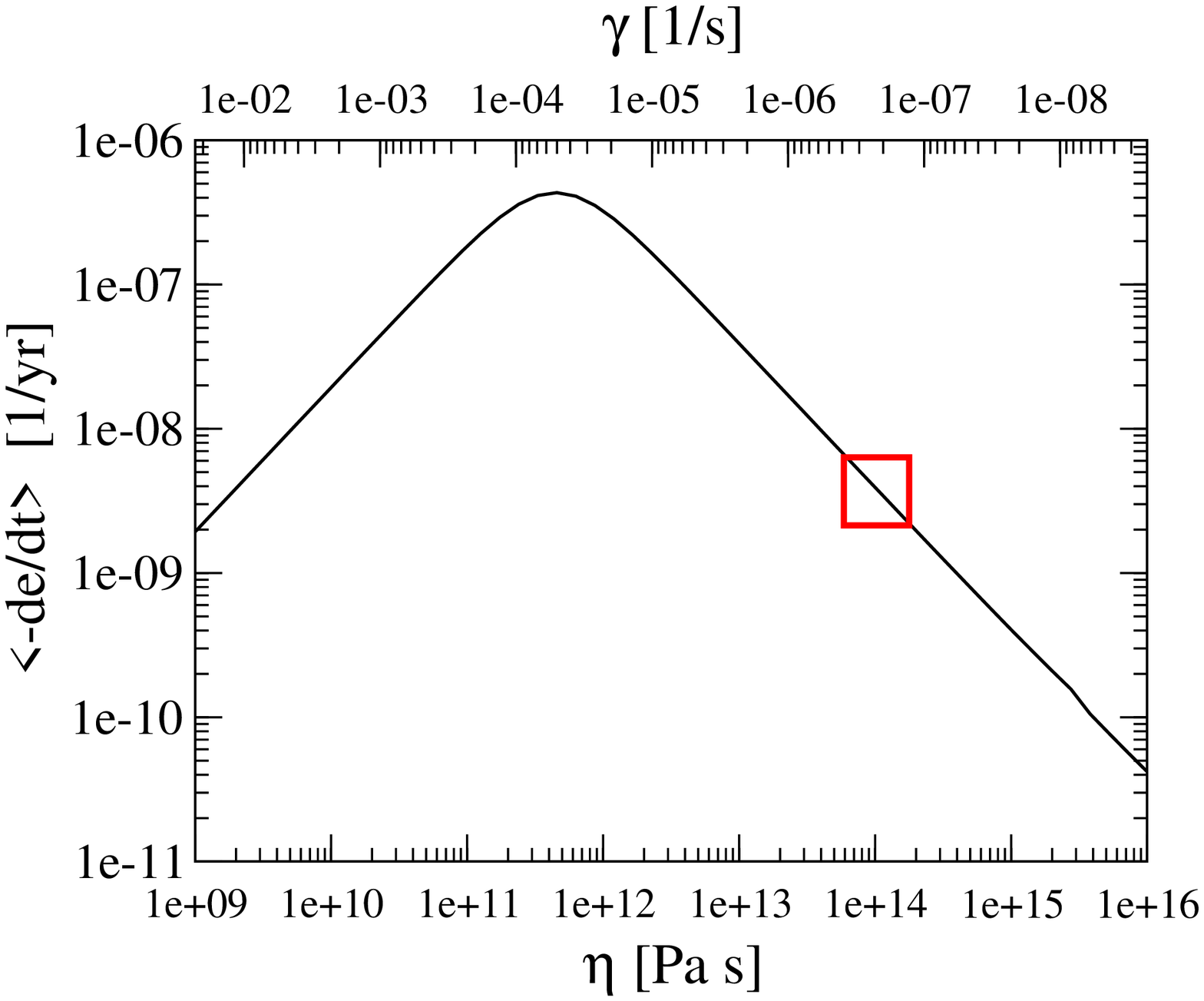}
\caption{Variation of the averages of $-\rm{d}a/\rm{d}t$ (\textit{Left}) and $-\rm{d}e/\rm{d}t$ (\textit{Right}) in function of the viscosity $\eta$ (or 
$\gamma$, see the top axis) for a body like Enceladus. In each plot the red box indicates the values corresponding to the actual range of the observed 
dissipation of Enceladus.}
\label{fig05}
\end{center}
\end{figure}

The variation of the semi-major axis due to the tidal deformations of the primary is given  by Eq. (\ref{eq:dadt1}). The results corresponding to a 
homogeneous Enceladus are shown in Fig. \ref{fig05} (Left). In that figure the average variation of $\rm{d}a/\rm{d}t$ is shown for a wide range of values of 
$\gamma$. The average has the same aspect as the dissipation law shown in Fig. \ref{fig04}. This similarity happens because, near the stationary solution, 
the dissipated energy comes almost totally from $\langle\dot{E}_{\rm orb}\rangle$ (the average variation of rotational energy $\langle\dot{E}_{\rm rot}\rangle$ is several orders of magnitude 
smaller). When $\gamma= 1.2-3.8 \times 10^{-7}\ {\rm s}^{-1}$, we obtain $\langle \dot{a}\rangle=-(0.4-1.2)\times 10^{-5}\ {\rm km/yr}$ (the red box in Fig. 
\ref{fig05} \textit{Left}).

It is important to emphasize that this is not the actual rate of change of the semi-major axis. This is only the part of it due to the tides on the 
satellite. To obtain the actual rate of change of ${a}$, it is necessary to consider also the part of it due to the tides raised on the planet, which is 
given by the same equations but where the variables are interchanged to express the dissipation on the planet instead of the satellite. In the case of 
Enceladus, the two effects appear to have similar orders of magnitude; according to Lainey et al. (2012), the variation of the semi-major axis due to the 
tides raised by Enceladus on Saturn is $4.2 \times 10^{-5}\ {\rm km/yr}$. The variation of the semi-major axis of Saturnian satellites is important for some 
theories of the formation of these satellites (see Lainey et al. 2012). The orbital variations due to the tides raised on the planet are more important than 
the variations due to the tides raised on the satellite and correspond to a present expansion of the satellite's orbit.

\subsection{Eccentricity}

The variation of the eccentricity is given by the corresponding Lagrange variational equation: 
\begin{equation}
 \dot{e}=-\frac{\sqrt{1-e^2}}{na^2e} \frac{\partial \cal R}{\partial \omega} + \frac{1-e^2}{na^2e} \frac{\partial \cal R}{\partial \ell},
\end{equation}
(see Brouwer and Clemence, 1961). This equation is equivalent to
\begin{equation}
\dot{e}=\frac{1-e^2}{e} \left(\frac{\dot{a}}{2a}-\frac{\dot{\cal L}}{\cal L}\right),
\label{eq:dedt1}
\end{equation}
where ${\cal L}= \frac{GMm}{na}\sqrt{1-e^2}$ is the orbital angular momentum. In order to use the averages given in the previous sections, we may introduce 
a change in this equation reminding that $\dot{\cal L}=M_z=-\dot{E}_{\rm rot}/\Omega$. Hence
\begin{equation}
\langle\dot{e}\rangle=\frac{1-e^2}{e} \left(\frac{\langle\dot{a}\rangle}{2a}+\frac{1}{\cal L}\left\langle\frac{\dot E_{\rm rot}}{\Omega}\right\rangle\right). 
\label{eq:edotav}
\end{equation}

The variation of the averaged $\rm{d}e/\rm{d}t$ as a function of the relaxation factor $\gamma$ is shown in Fig. \ref{fig05} (\textit{Right}). When 
$\gamma= 1.2-3.8 \times 10^{-7}\ {\rm s}^{-1}$, the averages are very small (less than $ 10^{-10}\ {\rm yr}^{-1}$). In general, the eccentricity variation 
may become important in long-term studies but, in the case of Enceladus, the effects of the almost 2:1 resonance between Enceladus and Dione produces a 
forced eccentricity of 0.00459 that must be considered (see Ferraz-Mello, 1985; Vienne and Duriez, 1995). We remind that the proper eccentricity of 
Enceladus is only 0.00012.

\subsection{Analytical approximations}

Proceeding similarly to the previous section, the average of the variation of the osculating semi-major 
axis due to the tide in the primary can be approximated, in the neighborhood of the synchronous rotation, as
\begin{equation}
\langle\dot{a}\rangle_{sync}=\frac{2a^2}{GMm}\langle\dot{E}_{\rm tot}\rangle_{sync} = -\frac{42R^2\overline{\epsilon}_\rho e^2}{5a}\frac{n^2\gamma}{n^2+\gamma^2},
\end{equation}
and the approximation of the average of the variation of the osculating eccentricity, in the neighborhood of the synchronous rotation, is
\begin{equation}
\langle\dot{e}\rangle_{sync}=\frac{1-e^2}{2ae}\langle\dot{a}\rangle_{sync}=-\frac{21R^2\overline{\epsilon}_\rho e(1-e^2)}{5a^2}\frac{n^2\gamma}{n^2+\gamma^2}.
\end{equation}

In both cases, the averages vanish when $e\rightarrow 0$. We may easily see that, when the quasi-synchronous rotation is assumed, both $\langle\rm{d}a/
\rm{d}t\rangle$ as $\langle\rm{d}e/\rm{d}t\rangle$ become proportional to $\langle\dot{E}_{\rm tot}\rangle_{sync}$.

\section{Extension to Mimas}

\begin{figure}[h]
\begin{center}
 \includegraphics[height=55mm,clip=]{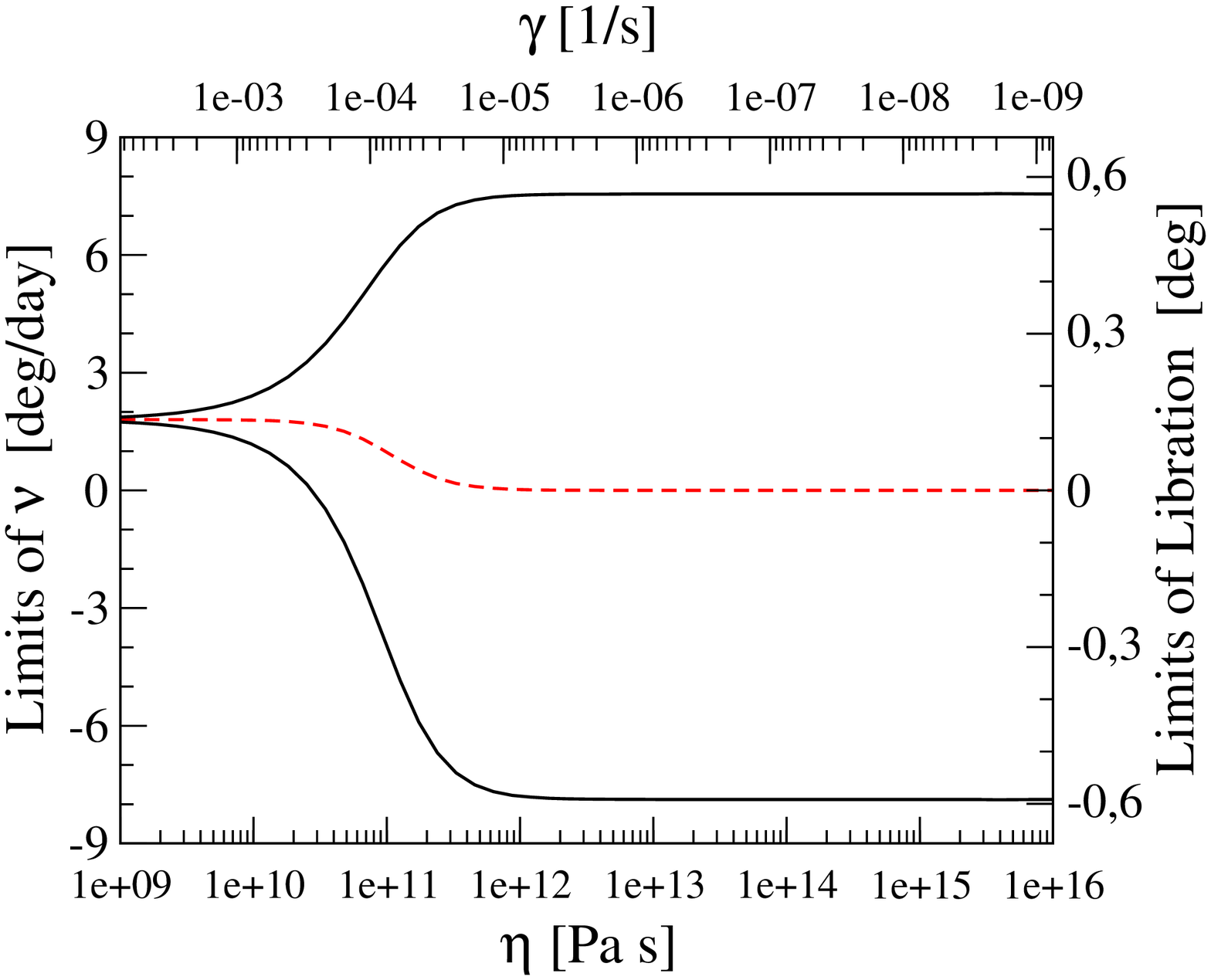}\ \ \ \ \ \ \ \ \ \ \ 
 \includegraphics[height=55mm,clip=]{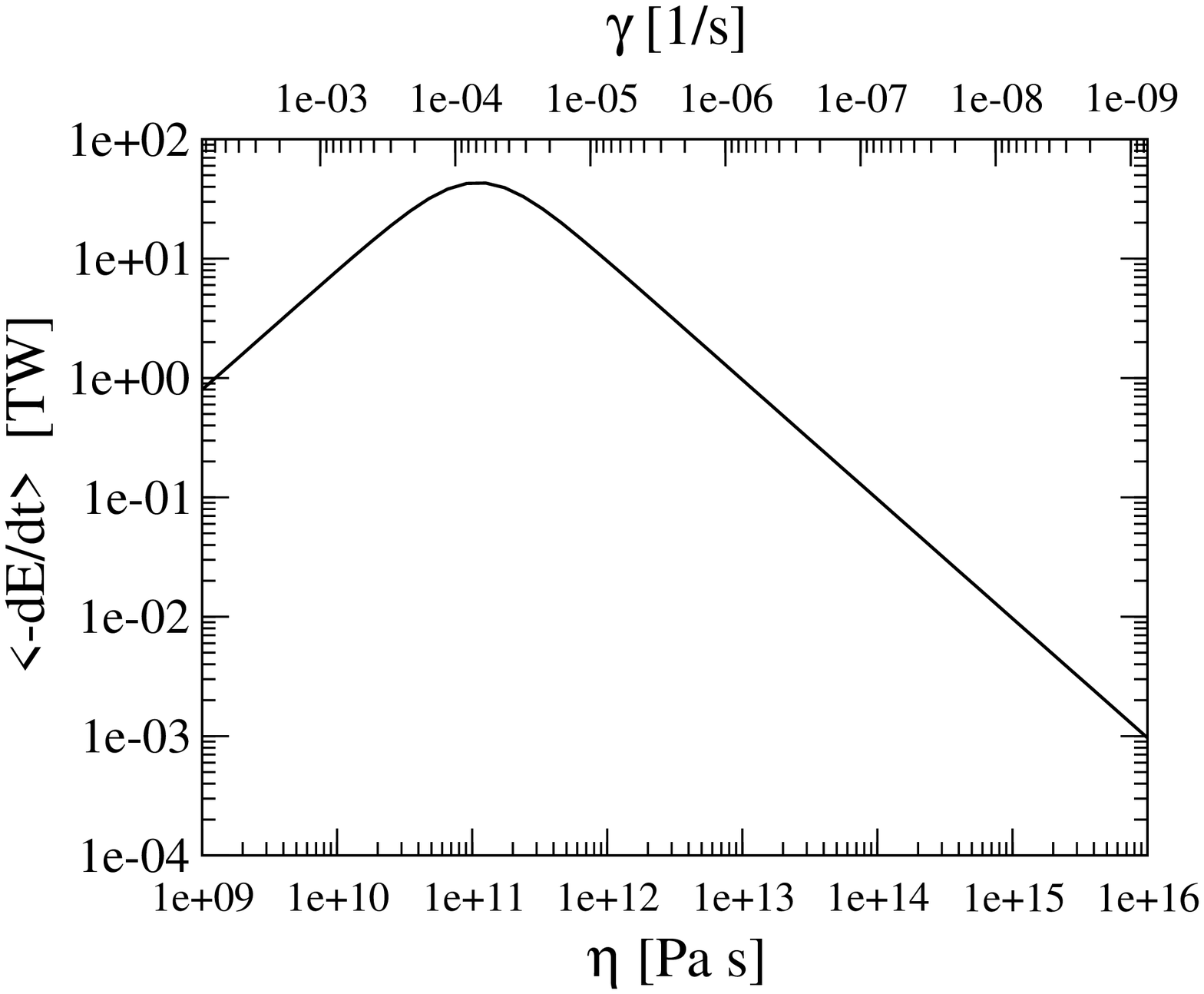}
\caption{\textit{Left:} Tidal forced oscillation of the semi-diurnal frequency (and the corresponding limits of the physical libration) of one body like 
Mimas in function of the viscosity $\eta$ and the relaxation factor $\gamma$ (shown on the upper axis). The black lines are the limits of the oscillation 
and the red dashed line is the mean value of $\nu$. \textit{Right:} Average of the net variation of the mechanical energy.}
\label{fig06}
\end{center}
\end{figure}

One challenge to every theory for the dissipation of the Enceladus is that it shall work also for the neighbor Mimas. Mainly, it must be coherent with the 
absence of tectonic activity in Mimas, evidence of a much smaller dissipation. The results obtained with the theory developed in this paper are shown in Fig. 
\ref{fig06}.

The first result concerns the physical libration. Measurements of control points on the surface of Mimas accumulated over seven years of Cassini's 
observations allowed Tajeddine et al. (2014) to determine the satellite's rotation state. They found several libration components including one short-period 
oscillation of $50.3 \pm 1\ {\rm arcmin}$. If we assume that these oscillations follow a harmonic law, we obtain, correspondingly, for the oscillation of 
the velocity of rotation: $5.6 \pm 0.1\ {\rm deg/day}$, and for the semi-diurnal frequency $11.2 \pm 0.2\ {\rm deg/day}$ (see Tables \ref{tab:data} and 
\ref{tab:resu}). The comparison of these results to those shown in Fig. \ref{fig06} (left) is that, like in the case of Enceladus, the observed value is 
larger than the tidal forced libration of Mimas (1.45 times) predicted by using a homogeneous body model. It is worth mentioning that this factor is not 
very different from the one obtained by Tajeddine et al. (2014) using models founded on the observed quadrupole moments of the gravitational potential of 
Mimas. In that case the relationship between the observed and the calculated amplitudes is 1.93 instead of 1.45 (compare with the 3.1 factor obtained in the 
case of Enceladus).

\begin{figure}[h]
\begin{center}
 \includegraphics[height=55mm,clip=]{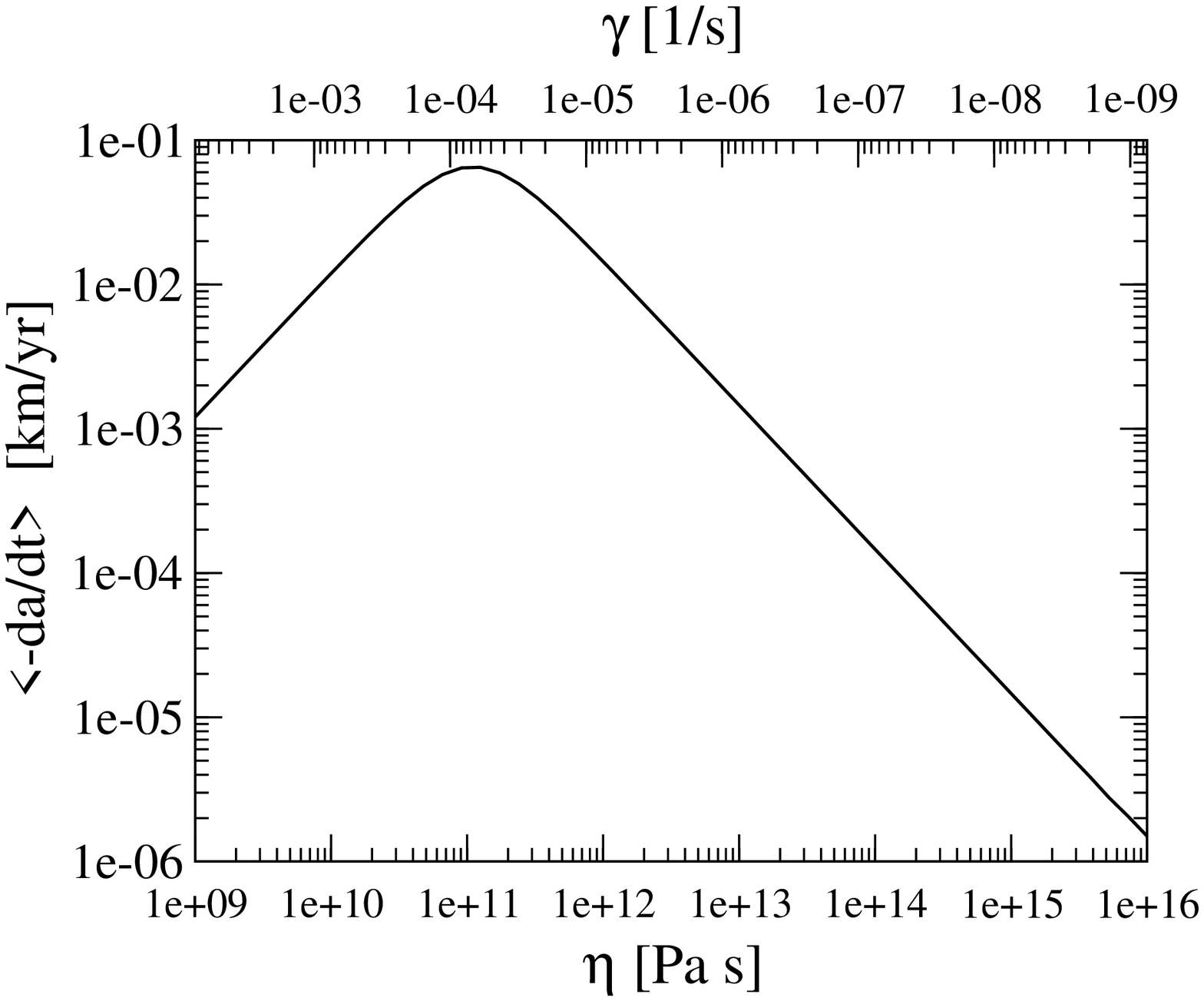}\ \ \ \ \ \ \ \ \ \ \ 
 \includegraphics[height=55mm,clip=]{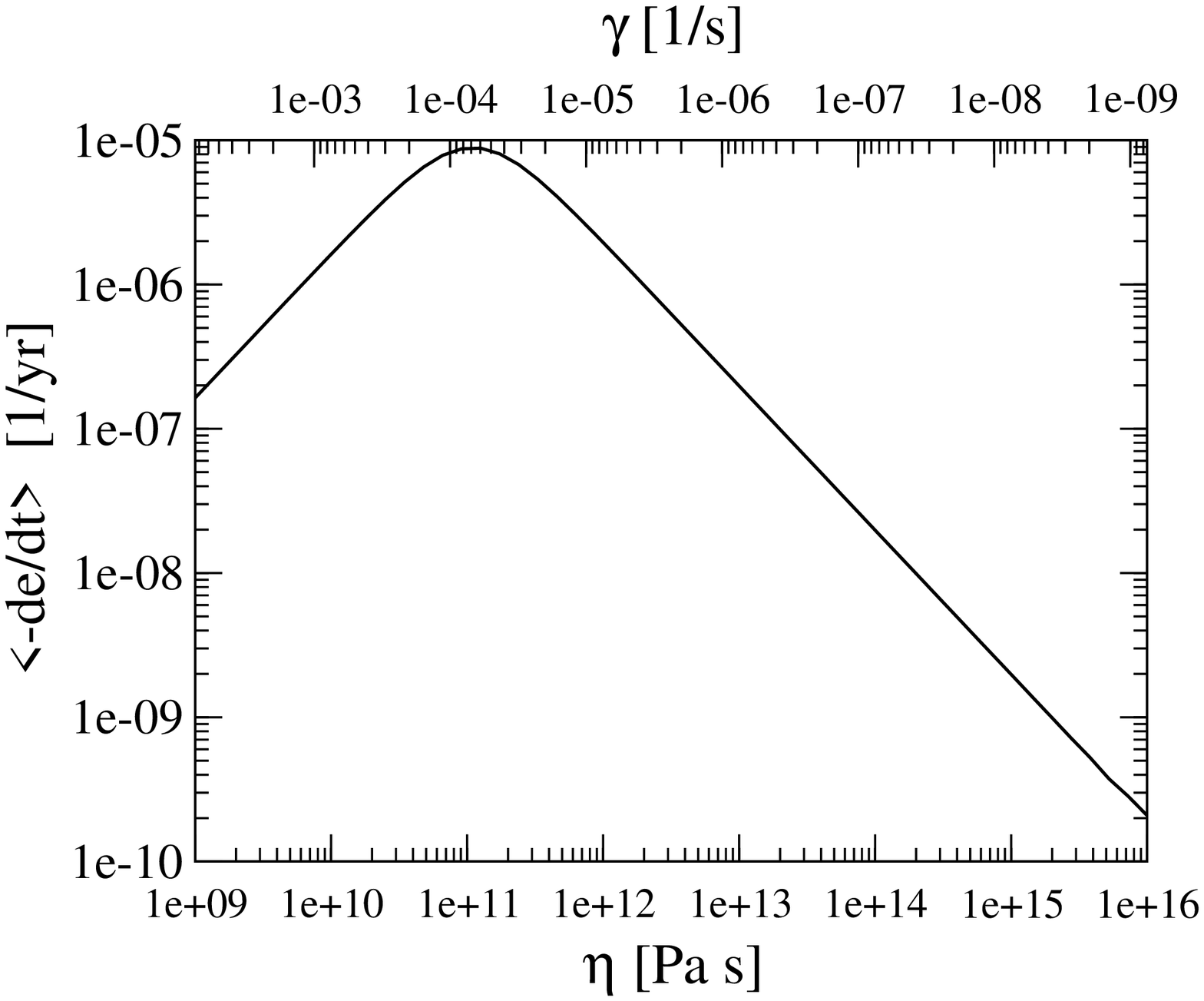}
\caption{Variation of the averages of $-\rm{d}a/\rm{d}t$ (\textit{Left}) and $-\rm{d}e/\rm{d}t$ (\textit{Right}) in function of the viscosity $\eta$ (or 
$\gamma$, see the top axis) for a body like Mimas.}
\label{fig07}
\end{center}
\end{figure}

The absence of current tectonic activity in Mimas is evidence of a small dissipation. Fig. \ref{fig06} (right) shows that a small dissipation indicates that 
the relaxation factor $\gamma$ of Mimas is much smaller than that of Enceladus. Let us consider it as $\gamma \sim 10^{-9}\ {\rm s}^{-1}$, that corresponds 
a dissipation smaller than 1 GW. The difference with the value found for Enceladus is surprising but it is consistent with the fact that the gravitational 
acceleration at the surface and the density are both much larger in Enceladus than in Mimas. Thus, the viscosity corresponding to the considered $\gamma$ is 
$\sim  10^{16}\ {\rm Pa\ s}$, which is a reasonable value.

For the sake of completeness, we may add the results on the contribution of the satellite tides to the orbital evolution of Mimas. With the relaxation 
factor considered above, we obtain $\langle\dot{a}_{sat}\rangle \sim -10^{-6}\ {\rm km/yr}$ and $\langle\dot{e}_{sat}\rangle \sim -10^{-10}\ {\rm yr}^{-1}$ 
(Fig. \ref{fig07}). These variations are much smaller than those due to the tides on the planet. Anyway, the current eccentricity of Mimas cannot be 
explained by tidal effects alone and is rather related to the crossing of mean-motion resonances in the past evolution of the satellite (Meyer and Wisdom, 
2008).

\begin{table}[t]
\caption{Observed libration and dissipation and results of the homogeneous model}
\begin{tabular}{lcc}
\hline 
 & {Enceladus} & {Mimas } \\
\hline
\\ 
Observed\\
Libration (deg) & $0.120\pm 0.014$&$0.838 \pm 0.002 $ \\
Dissipation (GW)& $5-16$& $-$ \\
Equatorial prolateness & $(2.0\pm 0.2)\times 10^{-2}$&$(5.5\pm 0.4)\times 10^{-2}$ \\
Polar oblateness & $(2.2\pm 0.1)\times 10^{-2}$&$(5.7\pm 0.2)\times 10^{-2}$ \\
\hline\\
Calculated\\
Libration (deg) & $0.039$ & $0.589  $ \\
Relaxation factor  (s$^{-1}$)& $1.2-3.8\times 10^{-7}$& $\sim 10^{-9}$ (*) \\
Viscosity (Pa s)& $0.6-1.9 \times 10^{14}$ & $\sim  10^{16}$ (*)\\
Semi-major axis variation (km/yr)& $-(0.6-1.8)\times 10^{-5}$& $\sim - 10^{-6}$ (*) \\
Eccentricity variation (yr$^{-1}$)& $-(1.9-6.0)\times 10^{-9}$ & $\sim- 10^{-10}$ (*)\\
Equatorial prolateness & $2.34\times 10^{-2}$ & $6.84\times 10^{-2}$ \\
Polar oblateness & $1.95\times 10^{-2}$ & $5.72\times 10^{-2}$ \\
\hline 
(*) Assuming dissipation $\sim 1 $ GW &&\\
\label{tab:resu}
\end{tabular}
\end{table}

\section{Conclusions}

In this paper we propose an improved approach of the original creep tide theory. Supported by the analytical solutions given in the previous papers 
(papers I and II), we assume that the tidally deformed body has a triaxial ellipsoidal shape, where the flattenings and orientation are unknown functions of 
the time to be determined. The creep tide equation allows us to find the differential equations that describe the time evolution of this ellipsoidal bulge 
and its orientation, resulting in a very simpler and compact approach.

The other main result of the present investigation is the dissipation law and its application to quasi-synchronous homogeneous bodies discussed in sections 
\ref{sec06} and \ref{sec07}. It is important to stress that the only hypothesis done in the theory is that the surface of the body permanently adjust itself 
to an equilibrium surface with speed given by the Newtonian creep law. No constitutive equation linking strain and stress is introduced at any point in the 
creep tide theory. All developments to reach the conclusion are the solution of the creep differential equation and the use of classical Physics to compute 
the force and torque acting on the external body due to the tidal deformation of the considered extended body. The observed dissipation law results directly 
from the above described first principles of Physics, with approximations but no additional ad-hoc hypotheses.

In the case of Enceladus, used in this paper as example of application, the estimations of the heat dissipated in the SPT (south polar terrain) area based 
on the observations with Cassini are in the range $5-16$ GW (\textit{cf} Howett et al. 2011; Spencer et al. 2013; Le Gall et al. 2017). In addition, a 
recent study by Kamata and Nimmo (2017) showed that a value about ten times higher than the old estimate of 1.1 GW is necessary if the ice shell is in 
thermal equilibrium. The given values correspond, using the creep tide theory, to a relaxation factor $\gamma = 1.2-3.8 \times 10^{-7}$ s$^{-1}$. The 
viscosity corresponding to this relaxation factor is $0.6-1.9 \times 10^{14}$ Pa s. It is of the same order as the value recently estimated by Efroimsky 
(2018) ($0.24 \times 10^{14}$ Pa s) and as the value adopted by Roberts and Nimmo (2008) for the viscosity of the ice shell ($10^{13}-10^{14}$ Pa s). This 
value is also close to the reference viscosity of water at 255 K ($10^{15}$ Pa s) adopted by B\u{e}hounkov\'a et al. (2012) in their modeling of the melting 
events at origin of the south-pole activity on Enceladus. More recent research carried out by \v{C}adek et al. (2019) demonstrates that the viscosity of ice 
at the melting temperature is equal to or higher than $3\times10^{14}\ {\rm Pa\ s}$, for the ice shell to remain stable.

For this range of values of $\gamma$, we can calculate the variation of the semi-major axis $\langle\dot{a}\rangle=-(0.4-1.3)\times 10^{-5}\ {\rm km/yr}$. 
For the variation of the eccentricity, we obtained a very small value $\langle \dot{e}\rangle=-(1.9-6.0)\times 10^{-9}\ {\rm yr}^{-1}$. However, in the case 
of Enceladus, the effects of the almost 2:1 resonance between Enceladus and Dione produces a forced eccentricity of 0.00459 that must be considered 
(see Ferraz-Mello, 1985; Vienne and Duriez, 1995).

In contrast with Enceladus, the absence of current tectonic activity in Mimas is evidence of a small dissipation. A dissipation $\sim 1$ GW, corresponds 
to a relaxation factor $\gamma \sim 10^{-9}\ {\rm s}^{-1}$ and a viscosity $\eta\sim 10^{16}\ {\rm Pa\ s}$, which are reasonable values. The difference with 
the values found for Enceladus is consistent with the fact that the gravitational acceleration at the surface and the density are both much larger in 
Enceladus than in Mimas. The different dissipations of Enceladus and Mimas may be simply associated with the fact that the outer layers of Enceladus have 
low viscosity (ice near the melting point) while Mimas, with no tectonic activity due to internal heating has a viscosity at least one order of magnitude 
larger (ice at temperatures well below the melting point).

Rough models of heat conduction considering the known conductivity of the ice at low temperatures show that the crust ability to convey heat 
produced in the interior is of some $10^{-2}$ W/m$^2$ and can be larger or smaller than the values discussed in this paper, depending on the ice crust width 
and properties. The temperature measurements of the surface of Enceladus (Spencer et al., 2006) show that most of the internally produced heat is flowing 
through the faults existing in the SPT and this confirms the inability of the existing crust ice to fully convey the produced heat. This behavior is a clue 
for a non-stationary process in which an increase in temperature means a decrease in the viscosity and a larger dissipation. In Enceladus, such a process 
may have been triggered by some transitory event enhancing the eccentricity of Enceladus and may have been progressing slowly, subsisting even after the 
eccentricity was damped to its current value. The transient increase of the eccentricity may have happened at any moment because of the small distance 
separating the inner satellites of Saturn (see Nakajima et al, 2018). 

Finally, measurements of control points on the surface of the satellites accumulated over seven years of Cassini's observations allowed Thomas et al. (2016) 
and Tajeddine et al. (2014) to determine their rotation state. They had found a libration of 0.120 $\pm$ 0.014 degrees for Enceladus and $50.3 \pm 1\ {\rm 
arcmin}$ for Mimas. Assuming that these oscillations follow a harmonic law, we obtain, correspondingly, for the oscillation of the semi-diurnal frequency 
$1.12 \pm 0.12\ {\rm  deg/day}$ for Enceladus and $11.2 \pm 0.2\ {\rm deg/day}$ for Mimas. These observed values are larger than the tidal forced libration 
predicted by using a homogeneous body model ($0.36\ {\rm  deg/day}$ and $7.7\ {\rm deg/day}$  for Enceladus and Mimas, respectively). These disagreements 
between theory and observations are mainly due to the assumed homogeneity of the satellites. Indeed, the Enceladus libration can be obtained using a 
multi-layered model (Folonier et al., in preparation), in which one liquid layer is assumed to exist between the crust and the core. All these results are 
summarized in Table \ref{tab:resu}.

\section* {Note added in proof}
The origin of the low dissipation value obtained in many papers using the classical models may be traced back to the use of Kelvin's formula for $k_2$ and 
arbitrarily fixed values for the rigidity leading to  $k_2 \le 0.002$. If more realistic values of $k_2$ (and $Q$), are used, as those determined by 
Choblet et al (2017, Supplement), the dissipation obtained with classical models is of the order of the observed values and coincide with the dissipation 
obtained in this paper when the viscosity is assumed to be that of melting ice. Efroimsky (2015, 2018) claims that for bodies of this kind, the rigidity 
plays virtually no role in tidal friction and $k_2$ is mainly defined by the viscosity of the body; thus, the use of Kelvin's formula in such cases is not 
correct. The tide theory used in the present paper also considers the viscosity rather than the rigidity and the comparison of the approximate formulas 
established in Section 8.3 of this paper with the corresponding ones in classical theories gives, for synchronous stiff bodies, 
$k_2/Q \simeq 1.5 \gamma / n = 0.75 wR/n\eta$ ($w$ is the specific weight).  This formula is virtually equivalent to the one  relating $k_2$ and the viscosity 
given by Efroimsky (2018), with only a difference in the numerical factor.

\begin{acknowledgement}
We thank Gwena\"el Bou\'e and Michael Efroimsky for their detailed reading of the original manuscript and for their enlightening suggestions. We also thank 
C. Beaug\'e and G.O. Gomes for several discussions. This investigation is funded by FAPESP, grants 2016/20189-9, 2014/13407-4, 2016/13750-6 and 2017/10072-0
and by the National Research Council, CNPq, grant 302742/2015-8. Preliminary results of this investigation were presented at the 9th Humboldt Colloquium on 
Celestial Mechanics, in Bad Hofgastein (Austria), March 2017.
\end{acknowledgement}

\section*{Appendix 1: Gravitational energy of an ellipsoid. Results after Ess\'en (2004)} \label{app:essen}

The gravitational energy of an ellipsoid of mass $m$ and semi-axes $a > b > c$ is explicitly given by 
\begin{equation}
E_{\rm int} = -\frac{3}{5} \frac{Gm^2}R X(\xi,\tau),
\end{equation}
(Ess\'en, 2004) where the Taylor expansion of $X$ around $\xi = 1, \tau = 0$ is
\begin{equation}
X(\xi,\tau)=1-\frac{4}{5}(\xi-1)^2 - \frac{4}{15}\tau^2 +\cdots,
\end{equation}
and $\xi$, $\tau$ are functions of the flattening such that
\begin{eqnarray}
a=(\xi+\tau)R; \ \ \ \ \ \ b=(\xi-\tau)R; \ \ \ \ \ \ c=(\xi^2-\tau^2)^{-1}R.
\end{eqnarray}
Hence, to the first-order,
\begin{eqnarray}
\tau=\half \mathcal{E}_\rho; \ \ \ \ \ \ \xi-1=\frac{1}{3}\mathcal{E}_z,
\end{eqnarray}
and
\beq
E_{\rm int} = -\frac{3}{5} \frac{Gm^2}R \left(1-\frac{1}{15}\mathcal{E}_\rho^2-\frac{4}{45}\mathcal{E}_z^2\right).
\label{eq:Wint}\endeq
The variation of the binding energy then is given by
\beq
\dot{E}_{\rm int} = \frac{3}{5} \frac{Gm^2}R \left(\frac{2}{15}\mathcal{E}_\rho\dot\mathcal{E}_\rho+\frac{8}{45}\mathcal{E}_z\dot\mathcal{E}_z\right).
\endeq
Eq. (\ref{eq:Wint}) agrees with the result of the direct integration showing that the difference between the binding gravitational energy of an ellipsoid 
and that of the corresponding sphere is of the order of the square of the flattenings $\mathcal{E}_\rho, \mathcal{E}_z$.

It is worth mentioning that the alternative formulation due to Neutsch (1979), does not agree neither with the results of Ess\'en (2004) nor with the 
results of a direct integration.

\section*{Appendix 2: Near-synchronous analytical approximation}\label{app02}

In order to find an analytical approximation to the instantaneous flattenings $\mathcal{E}_\rho,\mathcal{E}_z$, the rotation angle $\delta$ and the 
semi-diurnal frequency $\nu=2\Omega-2n$ in the quasi-synchronous attractor, let us first consider a Keplerian motion for the companion \tens{M}. Then, we 
consider the approximations:
\begin{eqnarray}
\left(\frac{a}{r}\right)^3  &\approx& 1+\frac{3e^2}{2}+3e\cos{\ell} \nonumber\\
n-\dot{\varphi}  &\approx& - 2ne\cos{\ell}.
\end{eqnarray}

Since, the rotation is quasi-synchronous, we may assume $\delta\ll 1$. Hence
\begin{equation}
\cos{2\delta} \approx 1-2\delta^2; \ \ \ \ \ \ \ \sin{2\delta} \approx 2\delta.
\end{equation}

Introducing these approximations into the creep tide and torque equations, (\ref{eq:system-creep}) and (\ref{eq:torque}) (using that $\nu=2\Omega-2n$, and 
$\dot\nu=2\dot\Omega$) and neglecting the term $\dot{C}\vec{\Omega}$, we obtain
\begin{eqnarray}
\dot{\nu} &=& -6\kappa n^2\left(1+\frac{3e^2}{2}+3e\cos{\ell}\right)\mathcal{E}_\rho\delta\nonumber\\
\dot{\delta}  &=&  \frac{\nu}{2}- 2ne\cos{\ell}-\frac{\gamma \overline{\epsilon}_\rho}{\mathcal{E}_\rho}\left(1+\frac{3e^2}{2}+3e\cos{\ell}\right) \delta \nonumber\\
\dot{\mathcal{E}}_\rho &=& \gamma \overline{\epsilon}_\rho\left(1+\frac{3e^2}{2}+3e\cos{\ell}\right)(1-2\delta^2)-\gamma \mathcal{E}_\rho    \nonumber\\
\dot{\mathcal{E}}_z &=& \frac{\gamma \overline{\epsilon}_\rho}{2}\left(1+\frac{3e^2}{2}+3e\cos{\ell}\right)+\gamma\overline{\epsilon}_z\frac{\Omega^2}{n^2}-\gamma\mathcal{E}_z,
\label{eq:sistema-ns}
\end{eqnarray}
where $\kappa=M/(M+m)$, and
\begin{equation}
\overline\epsilon_\rho=\frac{15MR_e^3}{4ma^3}; \ \ \ \ \ \ \ \overline\epsilon_z=\frac{5n^2R_e^3}{4Gm}.
\end{equation}
It is important to note that both $\overline\epsilon_\rho$ as $\overline\epsilon_z$ are constants.

Let us assume the particular solution
\begin{eqnarray}
 \nu              &=& B_0 + B_1\cos{\ell}+ B_2\sin{\ell}\nonumber\\
 \delta           &=& D_0 + D_1\cos{\ell}+ D_2\sin{\ell}\nonumber\\
 \mathcal{E}_\rho &=& E_0 + E_1\cos{\ell}+ E_2\sin{\ell}\nonumber\\
 \mathcal{E}_z    &=& Z_0 + Z_1\cos{\ell}+ Z_2\sin{\ell},
 \label{eq:part-solu}
\end{eqnarray}
the derivatives of which are
\begin{eqnarray}
 \dot{\nu}              &=& -nB_1\sin{\ell}+nB_2\cos{\ell}\nonumber\\
 \dot{\delta}           &=& -nD_1\sin{\ell}+nD_2\cos{\ell}\nonumber\\
 \dot{\mathcal{E}_\rho} &=& -nE_1\sin{\ell}+nE_2\cos{\ell}\nonumber\\
 \dot{\mathcal{E}_z}    &=& -nZ_1\sin{\ell}+nZ_2\cos{\ell}.
 \label{eq:part-solu2}
\end{eqnarray}

Finally, replacing (\ref{eq:part-solu}) and (\ref{eq:part-solu2}) into the system (\ref{eq:sistema-ns}), collecting the terms with same trigonometric 
argument and neglecting terms of higher order, we obtain 
\begin{eqnarray}
 B_0 &=& \frac{12n e^2}{1+p^2}\frac{1+\alpha p^2}{1+\alpha^2 p^2} \nonumber\\
 D_0 &=& \frac{3p e^2}{1+p^2}\frac{2+(1+\alpha)p^2}{1+\alpha^2p^2} \nonumber\\
 E_0 &=& \overline{\epsilon}_\rho\left(1+\frac{3e^2}{2}-\frac{4p^2e^2}{1+\alpha^2p^2}\right)\nonumber\\
 Z_0 &=& \frac{\overline{\epsilon}_\rho}{2}\left(1+\frac{3e^2}{2}\right)+\overline{\epsilon}_z\left(1+\frac{12e^2}{1+p^2}\frac{1+\alpha p^2}{1+\alpha^2p^2}+\frac{2(1-\alpha)^2p^2e^2}{1+\alpha^2p^2}\right),
\label{eq:X0}
\end{eqnarray}
and
\begin{eqnarray}
 \left(\begin{array}{c}  B_1\\
                         B_2 \end{array}\right) 
 &=& \displaystyle \frac{12\kappa\overline{\epsilon}_\rho np e}{1+\alpha^2p^2} \left(\begin{array}{c} -\alpha p\\
                                                                                          1 \end{array}\right) \nonumber\\
 \left(\begin{array}{c}  D_1\\
                         D_2 \end{array}\right) 
 &=& -\displaystyle \frac{2p e}{1+\alpha^2p^2} \left(\begin{array}{c} 1\\
                                                                      \alpha p \end{array}\right) \nonumber\\
 \left(\begin{array}{c}  E_1\\
                         E_2 \end{array}\right) 
 &=& \displaystyle\frac{3\overline{\epsilon}_\rho e}{1+p^2}  \left(\begin{array}{c} 1\\
                                                                                  p \end{array}\right) \nonumber\\
 \left(\begin{array}{c}  Z_1\\
                         Z_2 \end{array}\right) 
 &=& \displaystyle \frac{1.5\overline{\epsilon}_\rho e}{1+p^2} \left(\begin{array}{c} 1 -\frac{16\kappa p^2\overline{\epsilon}_z}{1+\alpha^2p^2}\\
                                                                                      p +\frac{8\kappa p(1-\alpha p^2)\overline{\epsilon}_z}{1+\alpha^2p^2} \end{array}\right),
\label{eq:Xi}
\end{eqnarray}
where $\alpha=1-3\kappa\overline{\epsilon}_\rho$ and $p=n/\gamma$. We mention that the mean values $B_0,D_0,E_0,Z_0$ are given to the second order in 
eccentricity while the amplitudes  $B_i,D_i,E_i,Z_i$ ($i\neq 0$) are given to the first order in eccentricity.

\begin{figure}[h]
\begin{center}
 \includegraphics[height=11cm,clip=]{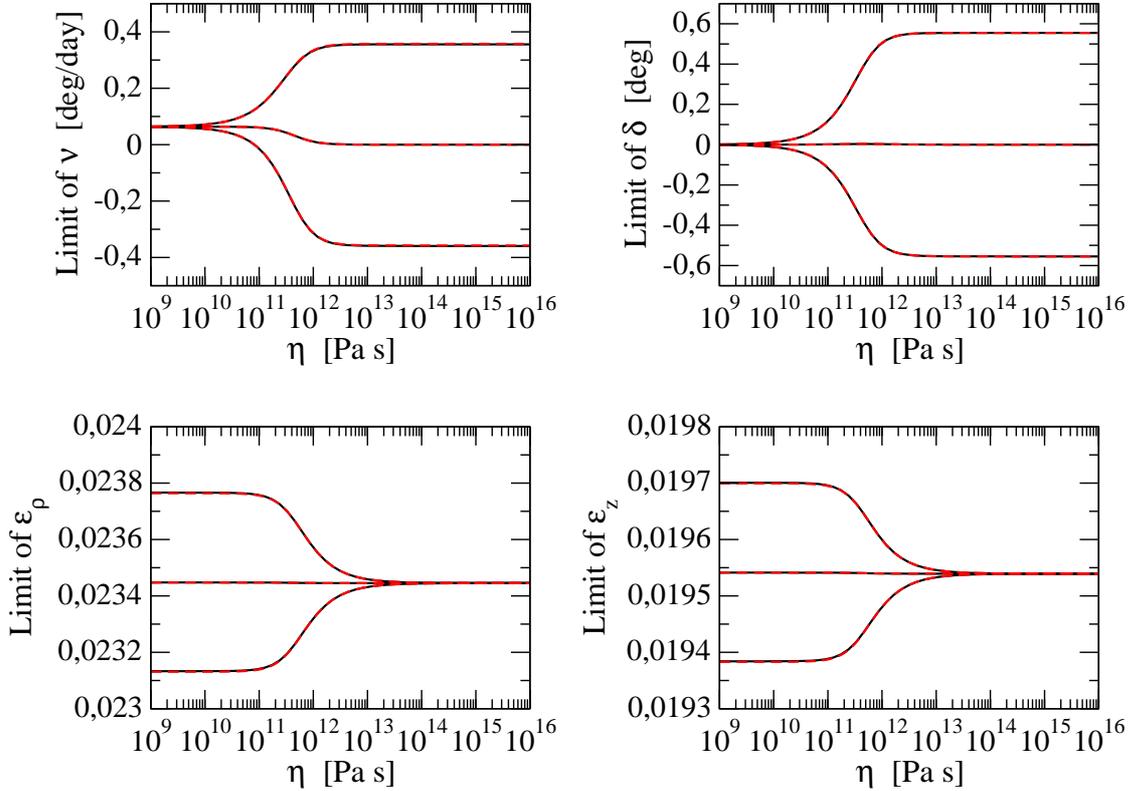}
\caption{Tidal forced oscillation of the semi-diurnal frequency $\nu$ (\textit{top left}), of the orientation angle of $\delta$ (\textit{top right}), of the 
equatorial prolateness $\mathcal{E}_\rho$ (\textit{bottom left}) and of the polar oblateness $\mathcal{E}_z$ (\textit{bottom right}) of one body like 
Enceladus in function of the viscosity $\eta$. The black solid lines are the limits of the oscillations and the mean values given by the numerical 
integration of Eqs. (\ref{eq:system-creep}) and (\ref{eq:torque}). The red dashed lines are the limits of the oscillations and the mean values given by the 
analytical approximation given by Eqs. (\ref{eq:part-solu}), (\ref{eq:X0}) and (\ref{eq:Xi}).}
\label{fig08}
\end{center}
\end{figure}

Figure \ref{fig08} shows the comparison of the near-synchronous attractors as given by the complete non-linear system defined by Eqs. (\ref{eq:system-creep}) 
and (\ref{eq:torque}) and the approximate analytical solution given above in the case of one body like Enceladus, in function of the viscosity $\eta$. The 
dashed red lines show the maximum, minimum and mean values of $\nu$, $\delta$, $\mathcal{E}_\rho$ and $\mathcal{E}_z$ (from \textit{top left} to 
\textit{bottom right}) given by the approximate solution, while the solid black lines show the maximum, minimum and the mean values of $\nu$, $\delta$, 
$\mathcal{E}_\rho$ and $\mathcal{E}_z$ when the complete non-linear system is integrated. The approximate solution is in excellent agreement with numerical 
integration of the equations. 

\section*{Corrections to Paper II (Ferraz-Mello, 2015)}
\begin{enumerate}
\item Typo in Eq. (2). The right definition is $\epsilon_z=1-\frac{c_e}{R_e}$.
\item In Eq. (16) the radial terms
$$\delta\zeta_{\rm rad} = - \frac {1}{3} R \sum_{k\in\Z}{\cal C}''_k \cos \overline\sigma''_k  \cos (k\ell-\overline\sigma''_k)$$
are missing. (N.B. They are torqueless and conservative and do not affect the results.)
\item Typo in Eq. (31). The argument should be $k\ell-\overline{\sigma}''_k$.
\item Mistake in Eq. (61). In the last line the arguments should be $v+k\ell-\overline{\sigma}''_k$ and $v-k\ell+\overline{\sigma}''_k$.
\item Mistake in Eqs. (62-68) The sign in front of the zonal part is wrong. The sign in front of ${\cal C}''_k$ in Eqs. (62-63) should be $+$, the sign in 
front of $kE_{0,k}^2$ in Eqs. (64-66) should be $-$ and the sign in front of $knE_{0,k}^2$ in Eq. (68) should be $+$.
\item Typo in Eq. (69). The sign in front of the right-hand side should be changed to $-$.
\item Mistakes in Eq. (70). The correct equation is
$$\dot{e}=-\frac{3GMR_e^2\overline\epsilon_\rho}{10na^5 e}\sum_{k\in \Z} E_{2,k}\cos \overline\sigma_k \sum_{j \in \Z}\Big(2\sqrt{1-e^2}-(2-k-j)(1-e^2)\Big) E_{2,k+j}\sin(j\ell+\overline\sigma_k)$$
$$\qquad -\frac{GMR_e^2}{10na^5 e} \sum_{k\in \Z} (\overline\epsilon_\rho E_{0,k}+ 2\delta_{0,k} \overline\epsilon_z) (1-e^2) \cos \overline\sigma''_k \sum_{j \in \Z} (k+j) E_{0,k+j}\sin(j\ell+\overline\sigma''_k).$$
\item Mistakes in Eq. (71). The correct equation is
$$\dot{e}=-\frac{3GMR_e^2\overline\epsilon_\rho}{20na^5 e}\sum_{k\in \Z} \Big(2\sqrt{1-e^2}-(2-k)(1-e^2)\Big)E_{2,k}^2\sin 2\overline\sigma_k$$
$$\qquad - \frac{GMR_e^2\overline\epsilon_\rho}{20na^5 e}\sum_{k\in \Z} (1-e^2)k E_{0,k}^2\sin 2\overline\sigma''_k.$$
\item Mistake in Eq. (B.6) (Online Supplement). The sign in front of $2\sqrt{1-e^2}E_{2,k}^{(5)}$ should be changed to $-$.
\end{enumerate}

{}
\vfill\eject

	\titlerunning{Tidal dissipation}
	\pagenumbering{arabic}
	\setcounter{section}{0}
	\setcounter{footnote}{0}
	\setcounter{equation}{0}  
	
	\bigskip
	\bigskip
	\begin{center}
		\LARGE{\textbf{Online Supplement}}
		\medskip \\
		\bigskip\medskip 
		\Large{\textbf{A self-consistent version of the creep tide theory \protect\\ valid for near-synchronous rotations}}\\
	\end{center}
	\medskip \bigskip
	$\,$\\
	\Large
	{Online Supplement to
		``Tidal synchronization of close-in satellites and exoplanets. III.  Tidal dissipation revisited and application to Enceladus" (H.A. Folonier, S. Ferraz-Mello and E. Andrade-Ines)}
	\medskip \bigskip
	\normalsize
	
	\section{Introduction}
	The creep equation, in the original version of the creep tide theory (Ferraz-Mello 2013 a.k.a. paper I), was solved assuming that, for short time spans, the 
	rotation of the primary body may be considered as uniform. One of the consequences of the theory was to show that, in near-synchronous motions, the tide 
	creates a short-period non-uniformity in the rotation of the primary, the physical libration (Ferraz-Mello 2015 a.k.a.paper II). Far from the spin-orbit 
	synchronization, the physical libration is just a minor perturbation of the motion and the uniform rotation may be considered as a good approximation. However, in 
	near-synchronous rotations, this is no longer the case and in order to have a self-consistent theory, it is necessary to consider \textit{ab initio } the physical 
	libration of the primary. This is done in this supplement\footnote{In the new version of the theory presented in the paper to which this supplement is 
		linked, a new formulation was adopted, where the rotation and the creep equations are solved simultaneously, and it is no longer necessary to make any 
		preliminary hypothesis on the rotational state of the primary.}.
	
	\section{The creep tide equation}\label{sec:creep}
	
	Let us consider one system formed by the extended body $\tens{m}$ (primary) and the mass point $\tens{M}$ (companion) and let $\textbf{r}$ be the 
	radius-vector of the companion in a system of reference centered in the primary. Since Darwin (1880), the problem of the tidal interaction of the two bodies 
	is split into two parts: First, we consider the deformation of the primary due to an external body (\tens{M}*) -- the body called \textit{Diana} by Darwin. 
	Then, we consider the disturbing potential due to this deformation on the companion, a mass point (\tens{M}) placed at a generic point of coordinates 
	$(r,\varphi,\theta) $. Eventually, we identify \tens{M} and \tens{M}*. The result is that the companion \tens{M} is moving in a time-dependent 
	potential field
	\begin{equation}
	U(\mathbf{r},\mathbf{r}^*),
	\end{equation}
	generated by \tens{m} in the point $\mathbf{r}$. The vector $\mathbf{r}^*$ is the radius 
	vector of \textit{Diana}. Even if \tens{M} and \textit{Diana} are physically the same body, the formulation should remain valid even if they were different 
	bodies. Thus, the force acting on \tens{M} due to the attraction of \tens{m} is
	\begin{equation}
	\textbf{f}= -M {\rm grad}_\mathbf{r} U(\mathbf{r},\mathbf{r}^*),
	\end{equation}
	where we included the subscript $\mathbf{r}$ to make clear that the derivatives in the evaluation of the gradient are done exclusively with respect to 
	$\mathbf r$.
	
	In the creep tide, the first half of Darwin's model - the tidal deformation of \tens{m} - is obtained by solving the Newtonian creep law
	\begin{equation}
	\dot{\zeta} = \gamma (\rho-\zeta),
	\label{eq:ansatz2}\end{equation}
	where $\gamma$ is the relaxation factor (see paper I), $\zeta=\zeta(\widehat\varphi,\widehat\theta,t)$ is the distance of the surface point of coordinates $\widehat\varphi$ (longitude) and $\widehat\theta$ 
	(co-latitude) to the center of gravity of the body, and $\rho=\rho(\widehat\varphi,\widehat\theta,t)$ is the surface of the static figure of equilibrium of 
	$\tens{m}$ under the gravitational attraction of \tens{M}. $\rho$ is approximated by a triaxial ellipsoid whose major axis is oriented towards  $\tens{M}$
	\begin{equation}
	\rho=R_e \left(1 + \half \epsilon_\rho \sin^2\widehat\theta \cos (2\widehat\varphi - 2\omega^* - 2v^*) - \epsilon_z \cos^2\widehat\theta \right),
	\end{equation}
	(see Tisserand, 1891; Chandrasekhar 1969; Folonier et al. 2015) where $\omega^*+v^*$ is the true longitude of \textit{Diana} in its equatorial orbit around 
	the primary ($\omega^*$ is the argument of the pericenter and $v^*$ is the true anomaly) and $\epsilon_\rho$ and $\epsilon_z$ are the equatorial prolateness 
	and polar oblateness, respectively. The angles are such that the major axis of the ellipsoid is always oriented towards \textit{Diana}. The right-hand side 
	is a known time function depending on  the longitude $\widehat\varphi$ and on the polar coordinates of the companion, $r^*$ and $v^*$. The radius vector of 
	\textit{Diana}, $r^*$, is introduced in the equation by the flattenings $\epsilon_\rho$ and $\epsilon_z$. 
	\begin{figure}
		\centerline{\hbox{
				\includegraphics[height=3.5cm,clip=]{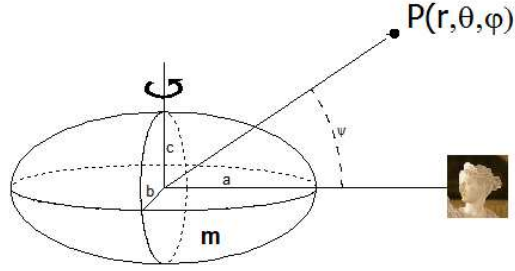}}}
		\caption{Darwin's model. The gravitational attraction of \textit {Diana} deforms the primary and the deformed primary attracts a mass placed at the point 
			\tens{P}. Afterwards, \tens{P} and Diana are identified one with the other.}
		\label{fig:diana}       
	\end{figure}
	
	If the Keplerian expansions of $r^*$ and $v^*$ are introduced (see papers I and II), the above equation becomes
	\begin{equation}\label{eq:rho}
	\rho= R  + R  \sum_{k\in\Z}\left( {\cal C}_k \sin^2\widehat\theta \cos \Theta_k +  {\cal C}''_k \left(\cos^2\widehat\theta-\frac{1}{3}\right) \cos \Theta''_k   \right),
	\end{equation}
	where 
	\begin{eqnarray}
	{\cal C}_k& \speq & \half \overline\epsilon_\rho E_{2,k} \nonumber\\
	{\cal C}''_k & \speq &-\half \overline\epsilon_\rho E_{0,k} - \delta_{0,k} \overline\epsilon_z,
	\end{eqnarray}
	and 
	\begin{eqnarray}
	\Theta_k &\speq & 2\widehat\varphi+(k-2)\ell^*-2\omega^*\nonumber\\
	\Theta''_k & \speq & k\ell^*. 
	\end{eqnarray} 
	$R$ is the mean radius of the primary (constant), $\ell^*$ is the mean anomaly, 
	\begin{equation}\overline\epsilon_\rho=
	\frac{15}{4}\left(\frac{M}{m}\right)\left(\frac{R_e}{a}\right)^3; \qquad
	\overline\epsilon_z=\epsilon_z-\half \epsilon_\rho,
	\end{equation}
	$\delta_{0,k}$ is the Kronecker delta function and $E_{p,q}(e)$ are the Cayley coefficients.
	
	Eqn. (\ref{eq:ansatz2}) is then a non-homogeneous ordinary differential equation of first order with constant coefficients. 
	
	\subsection{Integration. The free rotating case}\label{sec:zetafree}
	
	When the body is in free rotation (not trapped into a spin-orbit resonance), the angles $\Theta_k,  \Theta''_k $ are linear functions of the time and, after 
	integration, we obtain
	\begin{equation}\label{eq:zeta}
	\zeta=Ce^{-\gamma t} + R + \delta \zeta,
	\end{equation}
	where $C=C(\widehat\varphi,\widehat\theta)$ is an integration constant. The forced terms arising from the non-homogeneous part of the differential equation 
	are\footnote{The term 
		$$\delta\zeta_{\rm rad} = - \frac {1}{3} R \sum_{k\in\Z}{\cal C}''_k \cos \overline\sigma''_k  \cos (k\ell-\overline\sigma''_k),$$ 
		was discarded, in paper II, on the grounds of a heuristic reason: Radial terms in $\delta\zeta$ were considered as artifacts resulting from the model, since 
		the volume of the body must remain unchanged. Besides, the contribution of radial terms is torqueless and conservative. However, these terms correct the 
		non-conservation of the volume introduced by the zonal terms in the equation for $\zeta$ and, for the sake of completeness, they must be kept in the 
		expression of $\delta\zeta$.}\\
	
	\begin{eqnarray}
	\delta\zeta &=&  R \sum_{k\in\Z}\left({\cal C}_k\sin^2\widehat\theta \cos  \sigma_k \cos (\Theta_k- \sigma_k)+{\cal C}''_k \left(\cos^2\widehat\theta-\frac{1}{3}\right) \cos  \sigma''_k  \cos (\Theta''_k- \sigma''_k)\right), 
	\label{eq:dzeta}
	\end{eqnarray}
	where
	\begin{eqnarray}
	\tan \sigma_k&=&\frac{\dot\Theta_k}{\gamma};\qquad
	\cos \sigma_k=\frac{\gamma}{\sqrt{\dot\Theta_k^2+\gamma^2}};\qquad \ \
	\sin \sigma_k=\frac{\dot\Theta_k}{\sqrt{\dot\Theta_k^2+\gamma^2}};\nonumber\\
	\tan \sigma''_k&=&\frac{\dot\Theta''_k}{\gamma};\qquad
	\cos \sigma''_k=\frac{\gamma}{\sqrt{\dot\Theta_k^{\prime\prime 2}+\gamma^2}};\qquad
	\sin \sigma''_k=\frac{\dot\Theta''_k}{\sqrt {\dot\Theta_k^{\prime\prime 2}+\gamma^2}}. 
	\label{eq:sigmas}
	\end{eqnarray}
	
	\subsection{Integration. The spin-orbit or synchronous resonance.}\label{sec:zetasync}
	
	The synchronous attractor is given by
	\begin{equation}
	\nu \defeq 2(\Omega-n)=B_0+{\cal{A}} \cos(\ell^*+\ell_1),
	\end{equation}
	where the mean value $B_0$, the amplitude $\cal{A}$ and the phase $\ell_1$ are constants.
	
	In this case, the longitude of the the generic point on the surface of the primary is $\widehat\varphi = \varphi_F + \int_0^t \Omega dt $, that is
	\begin{equation}
	\widehat\varphi = \varphi_F + nt +\half B_0 t + \delta\varphi,
	\end{equation}
	where $\varphi_F$ is a constant and 
	\begin{equation}
	\delta\varphi = \frac{\cal{A}}{2n}\sin(\ell^*+\ell_1),
	\end{equation}
	is the periodic component in the variation of $\Omega$.
	
	Therefore, 
	\begin{equation}
	\Theta_k =  2\varphi_F + 2nt + B_0 t + 2\delta\varphi + (k-2)\ell^* -2\omega^*,
	\end{equation}
	and, expanding $\cos\Theta_k$ in the powers of $\delta\phi$,
	\begin{equation}
	\cos\Theta_k = \cos\overline\Theta_k - 2\delta\varphi  \sin \overline\Theta_k + \cdots,
	\end{equation}
	where
	\begin{equation}
	\overline\Theta_k =  2\varphi_F + B_0t + k\ell^* - 2\omega^*.
	\end{equation}
	If the definition of $\cal{A}$ is considered, we get
	\begin{equation}
	\cos\Theta_k = \cos\overline\Theta_k - \frac{\cal{A}}{n}\sin(\ell^*+\ell_1)  \sin \overline\Theta_k + \cdots,
	\end{equation}
	or
	\begin{equation}
	\cos\Theta_k = \cos\overline\Theta_k 
	- \frac{\cal{A}}{2n}
	\cos (\overline\Theta_{k-1} -\ell_1)  
	+ \frac{\cal{A}}{2n}
	\cos (\overline\Theta_{k+1} +\ell_1)  + \cdots .
	\end{equation}
	Thus, the original term in $\cos\Theta_k$ is decomposed into several terms whose arguments are linear functions of $t$ and the creep differential equation 
	can be integrated following the same simple rules as before. The forced terms then become
	\begin{eqnarray}
	\delta\zeta_{\rm res} &\speq&   R \sum_{k\in\Z} {\cal C}_k  \sin^2\widehat\theta \bigg[\cos  \overline\sigma_k \cos (\overline\Theta_k- \overline\sigma_k) - \frac{\cal{A}}{2n} \bigg(\cos  \overline\sigma_{k-1} \cos (\overline\Theta_{k-1}-\ell_1- \overline\sigma_{k-1})\nonumber\\
	&     &- \cos  \overline\sigma_{k+1} \cos (\overline\Theta_{k+1}+\ell_1- \overline\sigma_{k+1})\bigg) \bigg] + R \sum_{k\in\Z} {\cal C}''_k \left(\cos^2\widehat\theta-\frac{1}{3}\right) \cos  \sigma''_k  \cos (\Theta''_k- \sigma''_k),
	\end{eqnarray}
	where 
	\begin{equation}
	\tan \overline\sigma_k=\frac{\dot{\overline\Theta}_k}{\gamma} =
	\frac{B_0+kn}{\gamma} .
	\end{equation}
	
	\section{The disturbing potential $\delta U(\mathbf{r},\mathbf{r}^*)$}\label{sec:U}
	
	As in previous papers (papers I and II), the tidally deformed shape of the primary is composed of a series of added ellipsoidal bulges and the corrections to 
	the potential due to the tidal forces in one point of space whose spherical coordinates are $r,\varphi,\theta$ may be easily calculated. The resulting 
	disturbing potential is
	\begin{eqnarray}
	\delta U &\speq&  -\frac{GC}{2r^3}\sum_{k\in\Z} \bigg[3{\cal C}_k \sin^2\theta \Big[\cos  \overline\sigma_k \cos (2\varphi+(k-2)\ell^*-2\omega^*- 2\delta\varphi -\overline\sigma_k) \nonumber\\ 
	& & - \frac{\cal{A}}{2n} \Big(\cos  \overline\sigma_{k-1} \cos (2\varphi+(k-3)\ell^*-2\omega^*-\ell_1- 2\delta\varphi -\overline\sigma_{k-1})  \nonumber \\
	& & \qquad\qquad -  \cos  \overline\sigma_{k+1} \cos (2\varphi+(k-1)\ell^*-2\omega^*+\ell_1-2\delta\varphi - \overline\sigma_{k+1})\Big)\Big] \nonumber  \\
	& & +{\cal C}''_k \cos\sigma''_k (3\cos^2\theta -1) \cos(k\ell^*-\sigma''_k)\bigg],
	\label{eq:deltaU}
	\end{eqnarray}
	where $\ell^*$, $\omega^*$ and $e$ are, respectively, the mean anomaly, the argument of the pericenter and the eccentricity of the body source of the tidal 
	potential: \tens{M}* (i.e. \textit{Diana}) (to stress the origin of $\ell^*$, we  marked this anomaly with one star), and $G$ is the gravitational constant. 
	
	The above equation corresponds to the case where the rotation is following the near synchronous attractor; in the more usual case where the primary has a 
	free rotation, the equation is similar and can be obtained by just making $B_0={\cal{A}}=0$ in the above equation.
	
	We notice that the first terms of Eqn. (\ref{eq:deltaU}) are associated with sectorial components of the tidal displacement while the last term is 
	associated with zonal terms (terms independent of the longitudes). We emphasize that the correction due to the neglected radial term in the solution of the 
	creep equation in paper II (see footnote 2) is considered in Eqn. (\ref{eq:deltaU}) using the mean radius $R$ instead of the mean equatorial radius 
	$R_e$. The corresponding correction in the energy is of second order in the flattening.  
	
	The variable elements in $\delta U$ are $r$, $\varphi$, $\theta$ and $\ell^*$. Strictly speaking, there are other variable parameters in $\delta U$, as, for 
	instance, $n$, $e$, but their derivatives are first-order infinitesimals and thus, their contribution to the variation  of $\delta U$ will be of 
	second-order. The variation of $\Omega$ is important when $k=0$ and is considered in the given equations.
	
	We may substitute the variables $r$ and $\varphi$ by the Keplerian variables of the motion of \tens{M} around \tens{m}. Also, since we are only studying the 
	planar case, we adopt $\theta=\pi/2$ and $\varphi=v+\omega^*$ ($v$ is the true anomaly). If we introduce the two-body expansions via Cayley 
	coefficients\footnote{For practical formulas using the Cayley coefficients, see the Online Supplement linked to paper II.}, and for the sake of having a 
	simpler expression, we expand $\delta U$ in the powers of $\cal A$ and discard the quadratic terms, there follows:
	\begin{eqnarray}
	\delta U &\speq& -\frac{GC}{2a^3}\sum_{k\in\Z} \sum_{k+j\in\Z} \bigg[3{\cal C}_k E_{2,k+j} \Big[\cos  \overline\sigma_k \cos ((k+j-2)\ell +(2-k)\ell^* +\overline\sigma_k)  \nonumber\\ 
	& & -\frac{\cal{A}}{2n} \cos  \overline\sigma_{k} \Big(\cos ((k+j-2)\ell + (1-k)\ell^* - \ell_1 + \overline\sigma_k) \nonumber\\ 
	& & \qquad \qquad - \cos ((k+j-2)\ell + (3-k)\ell^* + \ell_1 + \overline\sigma_k) \Big)  \nonumber\\ 
	& & - \frac{\cal{A}}{2n} \Big(\cos  \overline\sigma_{k-1} \cos \big((k+j-2)\ell+(3-k)\ell^*+\ell_1+\overline\sigma_{k-1})  \nonumber\\ 
	& & \qquad\qquad -  \cos  \overline\sigma_{k+1} \cos ((k+j-2)\ell+(1-k)\ell^*-\ell_1 + \overline\sigma_{k+1})\Big)\Big] \nonumber\\
	& & -{\cal C}''_k E_{0,j+k}\cos\sigma''_k \cos((k+j)\ell - k\ell^* +\sigma''_k)\bigg],
	\label{eq:dU}
	\end{eqnarray}
	where the subscripts in the infinite summations were adjusted to mimic those used in paper II.
	
	\section{The rotation. Forced libration}\label{sec:Lib}
	
	When the time variation of the moment of inertia is neglected, the rotation of the body $\tens{m}$ is ruled by the differential equation $C\dot{\Omega}=M_z$ 
	where $M_z$ is the torque acting on the primary\footnote{$M_z$ is equal to the second component of the torque acting on the companion due to the tides of 
		${\tens m}$. To transform one into the other it is enough to change the sign twice. One because of the action-reaction principle and the other because in 
		the used coordinates, the co-latitude $\theta$ is reckoned downward. The minus sign appearing in Eqn. (\ref{eq:Mz}) comes from the construction of the 
		vector product $\mathbf{r}\times \mathbf{F}$.} and $C$ is the moment of inertia. The numerical  integration of this equation shows that the rotation is 
	tidally damped to a stationary solution which, when $\gamma$ is small, is a forced libration around a nearly synchronous center as in paper II. That is, a 
	periodic attractor. Using the definition of ${\cal C}_k$, we obtain:
	\begin{eqnarray}
	M_z &\speq& -M \displaystyle\frac{\partial\delta U}{\partial \varphi }  = -\frac{3GMC\overline\epsilon_\rho}{2r^3} \nonumber\\
	& & \times \sum_{k\in \Z}  E_{2,k} \bigg[\cos  \overline\sigma_k \sin (2\varphi+(k-2)\ell^*-2\omega^*- 2\delta\varphi -\overline\sigma_k)  \nonumber\\ 
	& & \quad \quad - \frac{\cal{A}}{2n} \Big(\cos  \overline\sigma_{k-1} \sin (2\varphi+(k-3)\ell^*-2\omega^*-\ell_1- 2\delta\varphi -\overline\sigma_{k-1}) \nonumber\\ 
	& & \qquad\qquad -  \cos  \overline\sigma_{k+1} \sin (2\varphi+(k-1)\ell^*-2\omega^*+\ell_1-2\delta\varphi - \overline\sigma_{k+1})\Big)\bigg].
	\label{eq:Mz} 
	\end{eqnarray}
	Hence, after introducing the two-body expansions, making $\ell^*=\ell$, dividing by the moment of inertia $C$ and expanding in the powers of $\cal A$, 
	(neglecting the terms of order ${\cal O(A}^2)$), we obtain:
	\begin{eqnarray}
	\dot\Omega &\speq& - \frac{3GM\overline\epsilon_\rho }{2a^3}\sum_{k\in\Z} \sum_{j+k\in\Z}E_{2,k} E_{2,k+j} \bigg[\cos  \overline\sigma_k \sin ( j \ell+ \overline\sigma_k)  \nonumber \\ 
	& & + \frac{\cal{A}}{2n} \cos \overline\sigma_k \Big( \sin((j+1)\ell+\ell_1+ \overline\sigma_k) -  \sin((j-1)\ell-\ell_1+ \overline\sigma_k )\Big) \nonumber\\
	& & - \frac{\cal{A}}{2n} \Big(\cos \overline\sigma_{k-1} \sin ((j+1)\ell+\ell_1+\overline\sigma_{k-1})-  \cos \overline\sigma_{k+1} \sin ((j-1)\ell-\ell_1 + \overline\sigma_{k+1})\Big)\bigg].
	\end{eqnarray}
	
	The average over one period is
	\begin{eqnarray}
	\langle\dot\Omega\rangle &\speq& - \frac{3GM\overline\epsilon_\rho }{2a^3}\sum_{k\in\Z} E_{2,k}  \bigg[E_{2,k}\cos  \overline\sigma_k \sin \overline\sigma_k \nonumber \\ 
	& & + \frac{\cal{A}}{2n} \cos \overline\sigma_k \Big( E_{2,k-1}\sin(\ell_1+ \overline\sigma_k) -  E_{2,k+1}\sin(-\ell_1+ \overline\sigma_k )\Big) \nonumber\\
	& & - \frac{\cal{A}}{2n} \Big(E_{2,k-1}\cos \overline\sigma_{k-1} \sin (\ell_1+\overline\sigma_{k-1})E_{2,k+1}\cos \overline\sigma_{k+1} \sin (-\ell_1 + \overline\sigma_{k+1})\Big)\bigg].
	\label{eq:avdotOmega}
	\end{eqnarray}
	
	For small amplitudes, the mean value $B_0$, the amplitude ${\cal A}$ and the phase $\ell_1$ of the variation of $\nu$ may be calculated using the 
	approximate formulas
	\begin{equation}
	B_0=\frac{12n\gamma^2e^2}{n^2+\gamma^2};\qquad
	{\cal A}=\frac{12n^2 e \overline\epsilon_\rho}{\sqrt{n^2+\gamma^2}}; \qquad 
	\ell_1=\arctan  \frac{\gamma}{n}.
	\end{equation}
	which are  simplifications of the formulas given by Folonier and Ferraz-Mello, 2017 (Online Supplement, Section B).
	
	\section{The energy balance}\label{sec:balance}
	
	\subsection{The orbital energy} \label{sec:Worb}
	
	In the adopted Darwinian model of Sections \ref{sec:creep}--\ref{sec:U}, $\delta U(\mathbf{r},\mathbf{r}^*)$ is the disturbing potential due to the 
	deformation of the mass (\tens{m}) acting on a generic point of coordinates $(r,\varphi,\theta) $ where the body \tens{M} is placed. Thus, actually \tens{M} 
	is moving in a time-dependent potential field and the energy of the system is
	\begin{equation}
	E_{\rm orb}=E_{\rm kin}+M U(\mathbf{r},\mathbf{r}^*),
	\end{equation}
	whose time variation is
	\begin{equation}
	\dot E_{\rm orb} = \dot{E}_{\rm kin} + Mn \frac{\partial U} {\partial \ell} + Mn \frac{\partial U} {\partial \ell^*},
	\end{equation}
	or 
	\begin{equation}
	\dot E_{\rm orb} = Mn \frac{\partial U} {\partial \ell^*}.
	\end{equation}
	The term $\dot{E}_{\rm kin}$ is canceled by an opposite term $-\dot{E}_{\rm kin}$ coming from the derivative with respect to $\ell$. The variations of the 
	slow variables $a,n$ are not considered in the differentiation because they do not give significant contributions at the order considered.  
	
	From Eqn. (\ref{eq:dU}), and making $\ell^*=\ell$, we obtain
	\begin{eqnarray}
	\dot{E}_{\rm orb} &\speq& \frac{3GMCn\overline\epsilon_\rho}{4a^3}\sum_{k\in \Z} \sum_{k+j\in \Z} \bigg[E_{2,k}E_{2,k+j} \Big[(2-k)\cos  \overline\sigma_k \sin (j\ell +\overline\sigma_k)  \nonumber\\ 
	& & -\frac{\cal{A}}{2n} \cos  \overline\sigma_{k} \Big((1-k)\sin (   (j-1)\ell  - \ell_1 + \overline\sigma_k)-(3-k)\sin ((j+1)\ell  + \ell_1 + \overline\sigma_k) \Big)  \nonumber \\ 
	& & -\frac{\cal{A}}{2n} \Big((3-k)\cos  \overline\sigma_{k-1} \sin \big((j+1)\ell +\ell_1+\overline\sigma_{k-1})-(1-k)\cos  \overline\sigma_{k+1} \sin( (j-1)\ell -\ell_1 + \overline\sigma_{k+1})\Big)\Big] \nonumber\\
	& & - \frac{k}{3}E_{0,k} E_{0,j+k}\cos\sigma''_k \sin( j\ell +\sigma''_k)\bigg].
	\end{eqnarray}

	\subsection{Rotational energy}\label{sec:Wrot}
	
	The variation of the rotational energy is given by the derivative of the kinetic energy: $C\Omega\dot\Omega$ where the moment of inertia $C$ is considered 
	as a constant. If we remember that, because of the forced libration, $\Omega$ is not constant, that is, 
	\begin{displaymath}
	\Omega=n \left(1+\frac{B_0}{2n}+\frac{\cal A}{2n} \cos(\ell+\ell_1)\right),
	\end{displaymath}
	expanding in the powers of $\cal A$, neglecting the terms of order ${\cal O(A}^2)$, and using the definition of ${\cal C}_k$, we obtain:
	\begin{eqnarray}
	\dot{E}_{\rm rot} &\speq& - \frac{3GMCn\overline\epsilon_\rho }{2a^3}\sum_{k\in\Z} \sum_{j+k\in\Z}E_{2,k} E_{2,k+j} \bigg[\cos  \overline\sigma_k \sin ( j \ell+ \overline\sigma_k)  \left(1+\frac{B_0}{2n}\right)\nonumber\\ 
	& & + \frac{\cal{A}}{4n} \cos \overline\sigma_k \Big( 3\sin((j+1)\ell+\ell_1+ \overline\sigma_k) -  \sin((j-1)\ell-\ell_1+ \overline\sigma_k )\Big) \nonumber\\
	& & - \frac{\cal{A}}{2n} \Big(\cos \overline\sigma_{k-1} \sin ((j+1)\ell+\ell_1+\overline\sigma_{k-1})- \cos \overline\sigma_{k+1} \sin ((j-1)\ell-\ell_1 + \overline\sigma_{k+1})\Big)\bigg].
	\end{eqnarray}
	
	\section{Dissipation}\label{sec:diss}
	
	\begin{figure}[]
		\begin{center}
			\includegraphics[scale=0.4]{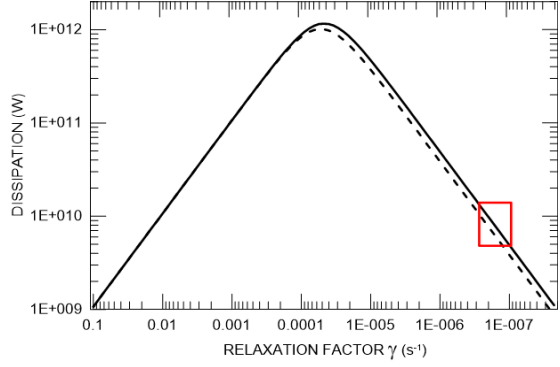}
			\caption{Dissipation curves when the body and orbital parameters used correspond to a homogeneous Enceladus (Eqs. 33 and 34).
				Solid line: Average of the net variation of the 
				mechanical energy obtained assuming for ${\cal A}$ the value corresponding to the observed physical libration of Enceladus. 
				Dashed 
				blue line: Average of the energy variation when making ${\cal A}=0$. The actual range of the observed dissipation values and the corresponding range for the 
				relaxation factor are shown by a red box.}
			\label{Q}
		\end{center}
	\end{figure}
	
	The variable characterizing the variation in a short time interval is the mean anomaly $\ell$ and the averaging operation is just $\frac{1}{2\pi}\int_0^{2\pi} 
	\dot{E}_{\rm tot} \D\ell$. The other elements are assumed to remain constant along one period. The averages of the variation of the orbital and rotational 
	energies over one orbital period are, respectively, 
	\begin{eqnarray}
	\langle\dot{E}_{\rm orb}\rangle &\speq& \frac{3GMCn\overline\epsilon_\rho}{4a^3}\sum_{k\in\Z} \bigg[E_{2,k} \Big[ E_{2,k}(2-k)\cos  \overline\sigma_k \sin \overline\sigma_k \nonumber\\ 
	& & +\frac{\cal{A}}{2n} \cos  \overline\sigma_{k} \Big(E_{2,k+1}(1-k)\sin(\ell_1-\overline\sigma_k)+ E_{2,k-1}(3-k)\sin ( \ell_1 + \overline\sigma_k) \Big)  \nonumber \\ 
	& & -\frac{\cal{A}}{2n} \Big(E_{2,k-1}(3-k)\cos  \overline\sigma_{k-1} \sin (\ell_1+\overline\sigma_{k-1})-E_{2,k+1}(1-k)\cos  \overline\sigma_{k+1}\sin( -\ell_1 + \overline\sigma_{k+1})\Big)\Big] \nonumber  \\
	& & +\frac{k}{3} E_{0,k}^2\cos\sigma''_k \sin \sigma''_k\bigg],
	\label{eq:Worbav}
	\end{eqnarray}
	and 
	\begin{eqnarray}
	\langle\dot{E}_{\rm rot}\rangle &\speq& - \frac{3GMCn\overline\epsilon_\rho }{2a^3}\sum_{k\in\Z} E_{2,k} \bigg[E_{2,k}\cos  \overline\sigma_k \sin \overline\sigma_k  \left(1+\frac{B_0}{2n}\right)\nonumber\\ 
	& & + \frac{\cal{A}}{4n} \cos \overline\sigma_k \Big(3 E_{2,k-1}\sin(\ell_1+ \overline\sigma_k) - E_{2,k+1}\sin(-\ell_1+ \overline\sigma_k) \Big) \nonumber\\
	& & - \frac{\cal{A}}{2n} \Big(E_{2,k-1}\cos \overline\sigma_{k-1}\sin (\ell_1+\overline\sigma_{k-1})-E_{2,k+1}\cos \overline\sigma_{k+1} \sin (-\ell_1 + \overline\sigma_{k+1})\Big)\bigg].
	\label{eq:Wrotav}
	\end{eqnarray}
	
	The energy variations obtained by using these equations coincide numerically with those obtained with the equations given in the main paper. It is worth 
	mentioning that, in the case of Enceladus, the dissipation obtained with the above equations taking the actual value of ${\cal A}$ given in this Supplement 
	is 27 percent larger than the results obtained when making ${\cal A}=0$, i.e., when neglecting the contribution of the physical libration (see Fig. \ref{Q}). 
	Thus, indeed, the physical libration contributes to increase the dissipation of a stiff body in near-synchronous rotation, but this contribution is not as 
	large as indicated by some previous calculations using the uniform rotation approximation in the solution for the creep equation. 
	
	The variations of the semi-major axis and eccentricity may be obtained in the same way as the equations given in Section 8 of the main paper.

	{}
	
\end{document}